\documentclass[preprint,12pt,authoryear]{elsarticle}

\usepackage{amsmath, amsthm, amssymb}
\usepackage{dsfont}

\usepackage[]{algorithm2e}
\usepackage{mdframed}

\newcommand{\R}{\mathbb{R}}
\newcommand{\El}{\mathcal{E}}
\newcommand{\e}{\varepsilon}
\newcommand{\n}{\nabla}
\newcommand{\Lop}{\mathcal{L}}
\newcommand{\1}{\mathds{1}}

\newcommand{\er}{v_{\mathrm{min}}}
\newcommand{\ef}{v_{\mathrm{max}}}
\newcommand{\Kstd}{K_{\rm{std}}}
\newcommand{\ds}{\displaystyle}
\newcommand{\vp}{\varphi}
\newcommand{\E}{\mathbb{E}}
\newcommand{\de}{\Delta}
\newcommand{\du}{\partial}

\newcommand{\mr}{\mathrm}

\newcommand{\nrSolvent}{62 }

\newtheorem{theorem}{Theorem}[section]
\newtheorem{remark}[theorem]{Remark}

\journal{Journal of Computational Physics}

\begin{document}

\begin{frontmatter}



\title{Estimating the speed-up of Adaptively Restrained Langevin Dynamics}


\author{Zofia Trstanova and Stephane Redon}

\address{
Univ. Grenoble Alpes, LJK, F-38000 Grenoble, France;
CNRS, LJK, F-38000 Grenoble, France; Inria, France}


\begin{abstract}

We consider Adaptively Restrained Langevin dynamics, in which the kinetic energy function vanishes for small velocities. Properly parameterized, this dynamics makes it possible to reduce the computational complexity of updating inter-particle forces, and to accelerate the computation of ergodic averages of molecular simulations. In this paper, we analyze the influence of the method parameters on the total achievable speed-up. In particular, we estimate both the algorithmic speed-up, resulting from incremental force updates, and the influence of the change of the dynamics on the asymptotic variance. This allows us to propose a practical strategy for the parametrization of the method. We validate these theoretical results by representative numerical experiments.
\end{abstract}


\end{frontmatter}



\section{Introduction}

One fundamental aim of molecular simulations is the computation of macroscopic quantities, typically through averages of functions of the variables of the system with respect to a given probability measure~$\mu$, mostly distributed according to the Boltzmann-Gibbs density. This corresponds to a model of a conservative system in contact with a heat bath at a fixed temperature. Numerically, the high-dimensional averages with respect to~$\mu$ are often approximated as ergodic\footnote{In this article we use the term ergodic for the convergence in infinite time, \textit{i.e.} the mathematical convergence (see~\cite{lelievre2016partial}).} averages over realizations of appropriate stochastic differential equations (SDEs) (see \cite{balian2006microphysics}).

In many applications, the dynamics is metastable, \emph{i.e.} the system remains trapped for a very long time in some region of the configuration space.
An acceleration of the sampling can be achieved by improving the exploration of the phase space (with variance reduction techniques such as importance sampling, see for instance \cite{lelievre2010free,lelievre2016partial}), increasing the time step by stabilizing the dynamics ( see for instance \cite{Matthews, fathi2015improving}), by changing the model as for example dissipative particle dynamics (see for instance \cite{espanol1995statistical}), or the implementation (parallelization, reduction of the computational complexity (see for instance \cite{bosson2012interactive})). Many methods are based on a modification of the dynamics. Since, very often, the interest lies in computing of average properties, sampling can be unbiased to retrieve averages with respect to the canonical distribution. In order to increase the time step size used in the simulation, some methods consider modifying the kinetic energy based on changing the mass matrix~(\cite{bennett1975mass,plechac2010implicit}). Another example is the Shadow-Hamiltonian Metropolis-Hastings method introduced by \cite{izaguirre2004shadow}, which consists in integrating the Hamiltonian dynamics according to the Shadow-Hamiltonian, which is preserved by the numerical scheme.

In order to propose a fair comparison of sampling methods, three factors should be taken into account: the asymptotic variance of time averages, the maximal admissible time step size in the discretization and the computational effort. 

In this work, we analyze the efficiency of a method based on a modified version of Langevin dynamics called ``Adaptively Restrained Particles Simulations'' (ARPS), first proposed in \cite{PRL-ARPS}. The main idea is to modify the kinetic energy function in order to freeze a number of particles at each time step and reduce the computational cost of updating inter-particles forces. In \cite{PRL-ARPS}, the kinetic energy is set to~0 when momenta are smaller than the restraining parameter $\er$, and is set to the standard, quadratic kinetic energy for momenta larger that the full-dynamics parameter $\ef$, with $\ef> \er\geq 0$. Thanks to this formulation, the computational complexity of the force update is reduced, because some particles do not move and hence, forces need not be updated. The associated gain can be quantified by an algorithmic speed-up factor~$S_{\rm a} \geq 1$. On the other hand, since the dynamics is modified, the asymptotic variance of time averages ~$\sigma^2_{\rm AR}$ given by the Central Limit Theorem, differs from the asymptotic variance ~$\sigma_{\rm std}^2$ of the standard Langevin dynamics. Intuitively, iterates are a priori more correlated, which may translate into an increase of the statistical error. The actual speed-up of the method in terms of wall-clock time is therefore an interplay between the algorithmic speed-up and the variances. A rigorous mathematical analysis of the ergodicity properties of this method was provided in \cite{trstanova2015errorAnalysisOfMLD}. Moreover, a perturbative regime study of the asymptotic variance suggests a linear behavior of the variance with respect to the parameters of the dynamics in some limiting regime.

Since the method is parameterized by two constants, it is not obvious how to choose these parameters in order to achieve an optimal speed-up. Of course, the algorithmic speed-up depends on the percentage of restrained particles. The percentage of restrained particles is a non-linear function of the parameters, hence it is not trivial how to  best choose their values. Our aim in this paper is to propose a strategy for choosing the parameters of the AR method.

The article is organized as follows: in Section \ref{Computation of macroscopic properties using MD simulations} we make a brief overview of sampling using Langevin dynamics and we recall common strategies for its discretization. In Section \ref{section AR-Langevin dynamics} we recall the definition of AR-Langevin dynamics proposed in \cite{PRL-ARPS} and the alternative definition of the AR-kinetic energy with better stability properties from \cite{NumericsInPreparation}. In Section \ref{section 2.3} we give a definition of speed-up and we introduce a formula for the total speed-up with the AR approach. In the next two sections we analyze how this formula depends on the parametrization: in Section \ref{section Algorithmic speed-up} we analyze the computational complexity of the method and we express the corresponding algorithmic speed-up. This part is followed by Section \ref{section Total speed-up}, in which we give a relation between the restraining parameters and the percentage of restrained particles, as well as an approach for obtaining the linear approximation of the variance with respect to the restraining parameters. By combining all the necessary parts, we propose a practical strategy for the parametrization of the method and we illustrate the theoretical results by numerical simulations in Section \ref{section numerical examples}.

\section{Modified Langevin dynamics}
\label{section 2}

In this section we recall the Langevin dynamics and the modified Langevin dynamics.

\subsection{Sampling from the canonical distribution using Langevin dynamics}
\label{Computation of macroscopic properties using MD simulations}

We consider a system of $N$ particles in a simulation box of the space dimension $D$ with periodic boundary conditions. The total dimension of the system is $d=N\times D$. We denote by $p=(p_1, \ldots, p_N)\in \R^d$ momenta of the particles and by $q=(q_1, \ldots, q_N)\in \mathcal{D}$ their positions in the box $ \mathcal{D}=(L\mathbb{T})^d$, where $\mathbb{T}=\R / \mathbb{Z}$ is the one-dimensional unit torus. We denote $\El:=\R^d\times\mathcal{D}$. The Hamiltonian of the system, which is the sum of the kinetic ($K$) and the potential energy ($V$), reads
\[
H(q,p)=K(p)+V(q)\, .
\]
Let us emphasize that we restrain ourselves to separable Hamiltonians.

The Langevin equations read:
\begin{equation}
  \label{modified Langevin}
  \left\{
  \begin{aligned}
    dq_t & = \n K(p_t) \, dt, \\
    dp_t & = -\n V(q_t) \, dt - \gamma \n K(p_t) \, dt + \sqrt{\frac{2\gamma}{\beta}} \, dW_t\,,
  \end{aligned}
\right.
\end{equation}
where $dW_t$ is a standard $d$-dimensional Wiener process, $\gamma>0$ is the friction constant and $\beta= (k_{\rm B} T)^{-1}>0$ is proportional to the inverse temperature. We refer the reader to \cite{lelievre2016partial} for an overview of mathematical properties of this dynamics. The invariant distribution (the Boltzmann distribution) is simply obtained as
\[
\mu(q,p)=Z^{-1}{\rm{e}}^{-\beta H(q,p)}, \quad \ds Z=\int_{\El}{\rm{e}}^{-\beta H(q,p)}dqdp \,,
 \]
where $Z$ is the normalization constant or the partition function. We use the notation $\mu_{\rm std}$ for the case of the standard kinetic energy function, and $\mu_{K}$ for a general kinetic energy function.
Langevin dynamics generate samples $(p_t, q_t)$ from the Boltzmann distribution $\mu$ which are used in computation of macroscopic properties. These correspond to expected values with respect to $\mu$ and can be approximated by ergodic averages of the trajectories $(p_t, q_t)$:
\begin{equation}
{\rm lim}_{t \rightarrow\infty}\hat{\vp}_t=\E_{\mu}(\vp)=\int_{\El} \vp d\mu, \quad \hat{\vp}_t=\frac{1}{t}\int_0^t\vp(q_s, p_s)ds\,.
\label{eq: ergodic averages}
\end{equation}
Note that for \textit{any} well-defined\footnote{We assume that $V$ belongs to $C^{\infty}(\mathcal{D},\R)$, and $K \in C^{\infty}(\R^d, \R)$ grows sufficiently fast at infinity in order to ensure that ${\rm{e}}^{-\beta K} \in L^1(\R^d)$.} $K$, due to the separability of the Hamiltonian, the marginal distributions in the position variable remain unchanged, since only the momenta marginals of the distribution are influenced by the modification of the kinetic energy, \textit{i.e.}
\[
\E_{\mu_{\rm{std}}}\left[\vp(q)\right]=\E_{\mu_{K}}\left[\vp(q)\right]\,.
\]

Since the dynamics \eqref{modified Langevin} cannot be integrated exactly its solutions are approximated by numerical integration (\cite{milstein2013stochastic}).
 Basically, there are two kinds of errors occurring in the estimation of $\E_{\mu}\left[\vp\right]$ by $\hat{\vp}_t$ from the numerical integration of equations \eqref{modified Langevin}: a statistical error, due to the finiteness of the time interval during which the sampling is performed; and a systematical error (bias) on the measure.

The statistical error for ergodic averages \eqref{eq: ergodic averages} is quantified by the Central Limit Theorem. The asymptotic variance associated with the estimator $\hat{\vp}_t$ reads
\[
\sigma^2=\lim_{t\rightarrow \infty}t \E_{\mu}\left[\left(\Pi{\vp}_t\right)^2\right]=2\int_0^{\infty}\E\left[\Pi\vp(q_s,p_s)\Pi\vp(q_0,p_0)\right]ds , \quad \Pi \vp:=\vp-\int_{\El}\vp d\mu\,.
\]
Similarly, for the discretized dynamics, with time step size $\de t=T/N_{{\rm iter}}$, we denote the estimator
\[
\hat{\vp}_{N_{\rm{iter}, \de t}}=\frac{1}{N_{\rm iter}}\sum_{i=0}^{N_{\rm iter}-1}\vp(q^n, p^n).
\]
 If the discretized dynamics is geometrically ergodic with an invariant measure $\mu_{\de t}$, a Central Limit Theorem holds true and the variance of the discretized process is given by (see \cite{MT93, lelievre2016partial})
\begin{equation}
\begin{aligned}
\ds \sigma^2_{\de t} &= \lim_{N_{\rm iter } \rightarrow \infty}N_{\rm{iter}}\rm{Var}_{\mu_{\de t}}\left(\hat{\vp}_{N_{\rm{iter}, \de t}}\right) \\
&= \E_{\mu_{\de t}}\left(\left[\Pi_{\de t}\vp \right]^2\right)+2\sum_{n=1}^{\infty}\E_{\mu_{\de t}}\left[\Pi_{\de t}\vp(q^n,p^n)\Pi_{\de t}\vp(q^0,p^0)\right] \\
&=\int_{\El}\Pi_{\de t} \vp\left[\left(2\left({\rm Id}-P_{\de t}\right)^{-1}-{ \rm Id}\right) \Pi_{\de t} \vp \right] \ d\mu_{\de t}\,
\end{aligned}
\label{eq: discrete variance}
\end{equation}
where
\[
\Pi_{\de t} \vp:=\vp-\int_{\El} \vp d\mu_{\de t}.
\]
In other words, for $N_{\rm{iter}}$ simulation steps, the statistical error is of order $\e:=\frac{\sigma_{ \de t}}{\sqrt{N_{\rm{iter}}}}$. The variance of the discretized process converges to the variance of the continuous process $\sigma^2$ as $\de t$ tends to $0$, i.e. $\de t\sigma^2_{ \de t}\rightarrow \sigma^2$ as $\de t \rightarrow 0$ (see \cite{lelievre2016partial}).

There are many possible ways to discretize (\ref{modified Langevin}), see for instance \cite{Matthews, mattingly2002ergodicity, Kopec2014} for a precise analysis of the properties of discretization schemes of the Langevin dynamics based on a splitting. A standard choice for the discretization of \eqref{modified Langevin} is a numerical scheme of second order on the averages with respect to the time step size. It is possible to design higher order schemes, however  they include at least double evaluation of the forces, which is not favorable due to the system size. Usually, the numerical schemes are constructed through a splitting of the generator of the Langevin dynamics $\Lop = A+B+ O$ with
\[
A:=\n K(p) \cdot \n_q, \qquad B:=-\n V(q)\cdot \n_p, \qquad O:= -\gamma\n K(p)\cdot \n_p + \frac{\gamma}{\beta }\de_p \,.
\]
For instance, the first order splitting (Lie-Trotter) gives the following scheme:
\begin{equation}
\ds
\left\{
\begin{aligned}
\ds\tilde{p}&= p^n-\n V(q^{n})\de t ,\\
\ds q^{n+1} &= q^n+\n K(\tilde{p}) \de t ,\\
\ds p^{n+1} &= \tilde{p}-\gamma \n K(\tilde{p})\de t + \sqrt{\frac{2\gamma \de t}{\beta}}G^{n}\,,
\end{aligned}
\right.
\label{discr modified Langevin first order}
\end{equation}
where $G^n$ are independent and identically distributed (i.i.d.) standard $d$-dimensional Gaussian random variables. This is the so-called BAO scheme. The name is motivated by the fact that the transition kernel reads $P^{\rm BAO}_{\de t}\vp=P^{\rm B}_{\de t}P^{\rm A}_{\de t}P^{\rm O}_{\de t}\vp$ where $P_{\de t}^A$ (respectively $P_{\de t}^B,P_{\de t}^O$) is the transition operator associated with the splitting step.
We refer to \cite{NumericsInPreparation} for a construction of second order discretization schemes in the case of a general kinetic energy.

\begin{remark}
The Lie-Trotter and the Strang splitting each give six possible numerical schemes according to the order of the operators A,B,O (\cite{Matthews}).
Due to the high dimensionality of the system, the bottleneck of the computational complexity is the computation of the interactions between the particles, which must be done after each update of the positions (after action A). In addition, a non-negligible computational effort involves generation of random numbers in O. Therefore, schemes which include as few actions of A and O as possible should be preferred for a lower computational complexity.
\end{remark}

\subsection{AR-Langevin dynamics}
\label{section AR-Langevin dynamics}

In the usual setting, the kinetic energy function of system of $N$ particles is a sum of kinetic energies of each particle, which are quadratic functions of momenta:
\[
\Kstd(p)=\sum_{i=1}^Nk(p_i)\,,\quad k(p_i)=\frac{p_i^2}{2m_i},
\]
 where $p_i\in \R^D$ is the momentum vector of the particle $i$ with mass $m_i$.

\begin{figure}[h]
	\centering
		\includegraphics[width=0.80\textwidth]{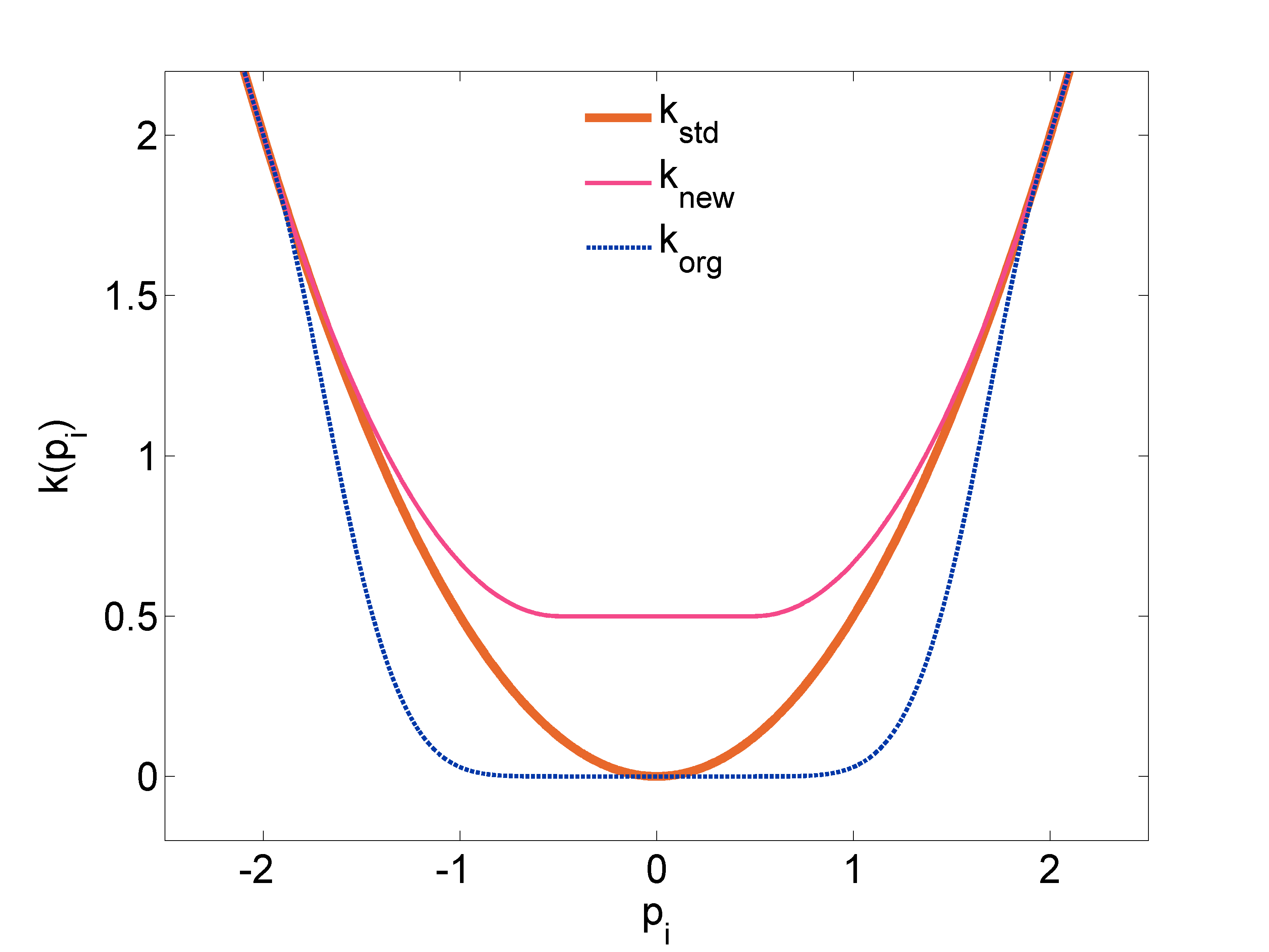}
		\caption{\textit{Comparison of the AR-kinetic energy functions for one particle ($p=p_0, N=1$)~(\cite{NumericsInPreparation})}: $k_{\rm org}$ is the original definition from \cite{PRL-ARPS}, $k_{\rm new}$ corresponds to Equation \eqref{def new arps kinetic energy} and $k_{\rm std}$ is the standard one. }
	\label{fig:modifiedHamiltonian_parametrization}
	\end{figure}

\begin{figure}[h]
	\centering
		\includegraphics[width=0.80\textwidth]{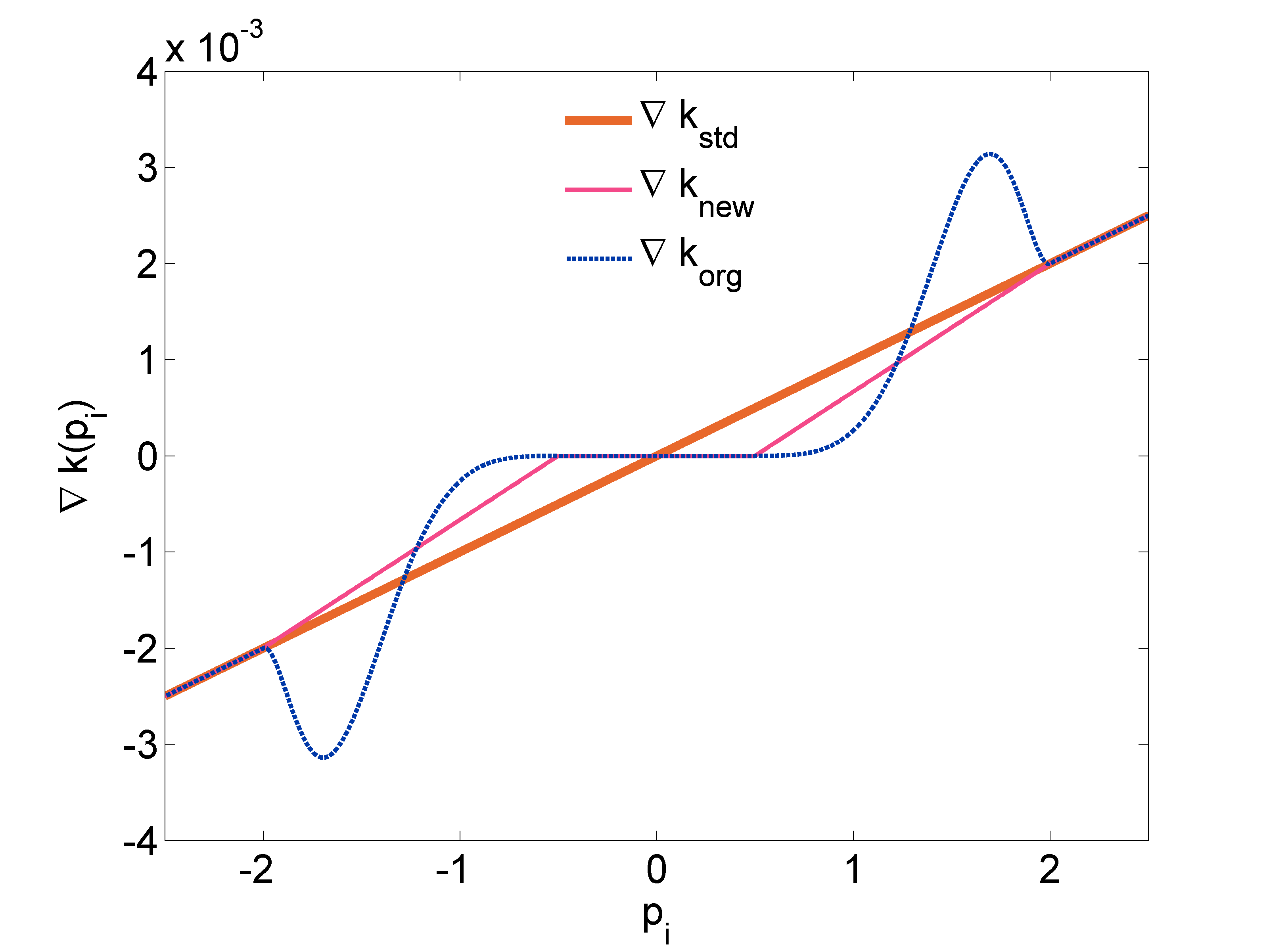}
		\caption{\textit{Gradient interpolation of the kinetic energy ($k_{\rm new}$) versus function interpolation ($k_{\rm org}$)}, see also Figure~\ref{fig:modifiedHamiltonian_parametrization}. Both gradients are vanishing around zero, which allows to freeze the positions of the particles. However, the gradient interpolation has been proven to be more numerically stable.}
\label{fig:diff_modifiedHamiltonian_parametrization}
	\end{figure}

AR-Langevin dynamics, proposed in \cite{PRL-ARPS}, is based on a modified kinetic energy function $K$ that is defined as the sum of the kinetic energies of the individual particles, which are parametrized by two constants $\ef~>~\er~\geq~0$. In \cite{PRL-ARPS}, the AR-kinetic energy was designed such that it vanishes for values smaller than the restraining parameter $\er$, and are equal to the standard kinetic energy for values bigger than the full dynamics parameter $ \ef$. The main idea is that the derivative $\du_{p_i} k$ of such function vanishes in this case too, and the position of the particle does not change between the two integration steps, i.e. $q_i^{n+1}=q_i^{n}, i\in I_{R}$, with $I_R$ being the set of indices of the restrained particles (see also Equation~\eqref{discr modified Langevin first order} for the computation of $q^{n+1}$). The transition between the restraining and the full-dynamics region is performed with an interpolation spline, which ensures the regularity of the kinetic energy\footnote{The choice was $k\in C^2(\R^D)$ in \cite{PRL-ARPS}.}. Still, the derivatives of $k$ have large values in the transition region, which might cause numerical instabilities. However, the necessary condition for the particle to remain at the same position between two time steps is that the first derivative of the kinetic energy function vanishes for some values of momenta in every space direction. In \cite{NumericsInPreparation} an alternative definition of the AR-kinetic energy function with this property was introduced, based on the vanishing velocities. The AR-kinetic energy is defined starting from an interpolation of its first derivative, such that it vanishes around zero and takes values of $p_i/m_i$ outside the restraining region (see Figure~\ref{fig:diff_modifiedHamiltonian_parametrization} for an illustration). The order of the interpolation spline can be chosen as high as necessary. However in this article we use only linear interpolation. The modified kinetic energy is then obtained by piecewise integration (see Figure~\ref{fig:modifiedHamiltonian_parametrization} for an illustration):
 \begin{equation}
\begin{aligned}
\displaystyle K(p)=\sum_{i=1}^d k(p_i)
 \quad\text{ where }\quad
\ds
	k(p_i)=
	\left\{
	\begin{aligned}
S_{\er,\ef}& \quad \text{ for } &  \frac{\left|{p_{i}}\right|}{m_i}\leq \er,\\
s\left(p_i\right) & \quad \text{ for } & \frac{\left|{p_{i}}\right|}{m_i}\in\left[\er, \ef\right],\\
		 \frac{p_i^2}{2m_i}&\quad  \text{ for } &  \frac{\left|{p_{i}}\right|}{m_i}\geq \ef\, ,
			\end{aligned}
	\right.
	\end{aligned}
	\label{def new arps kinetic energy}
	\end{equation}
	for $i=1,\ldots d=N\times D$, and where $s$ is the integrated interpolation spline and $S_{\er,\ef}$ is an integration constant, which corresponds to the minimal kinetic energy value of the particle\footnote{We would like to emphasize that this constant does not appear in the Langevin equations, only in the momenta marginal of the invariant measure.}. Note that the total kinetic energy is the sum of individual kinetic energies of each particle in every space direction. 

AR dynamics accelerate sampling by exploiting information about the kinetic energy of particles. More precisely, a particle is called \textsl{restrained} if it has the absolute value of each component of its momentum smaller than the restraining threshold $\er$. All other particles are defined as \textsl{active} particles. Note that, during the simulation, particles are switching between these states. Also, the average occupation of the active or restrained state only depends on the restraining parameters $\er$ and $\ef$.

	 Since the momenta of individual particles are independent from each other under the canonical measure, the parameters $\er$ and $\ef$ could in fact be different for each particle, to either focus calculations on a specific part of the particle system or to adjust the scaling of parameters according to the mass of the particle.

\section{Estimating the speed-up }
\label{section 2.3}
 In this section we introduce a framework for the complexity analysis of the AR dynamics in the case of pair-wise interactions, which are the most common interactions in numerous applications. Note that the discussion below can be easily generalized to interactions present in classical force-fields. The force acting on each particle $i$ is a sum of interactions with all other particles.

The information about the state of the particle allows us to lower the computational cost of the computation of pair-wise interactions between the particles. We consider the potential
\[
V(q)=\sum_{\substack{j=1 \\ i\neq j}}^N v(r_{ij})
\]
and the force acting on the particle $i$ which is given by
\begin{equation}
f_i(q):=-\du_{q_i} V(q)=-\sum_{\substack{j=1 \\ i\neq j}}^N v^{\prime}(r_{ij})\frac{q_i-q_j}{r_{ij}}\,,\quad r_{ij}=\left|q_i-q_j\right|\,.
\label{force on one particle}
\end{equation}
The change of the force between two time steps only depends on active particles that have moved since the last time step with respect to this particle. This allows us to avoid the computation of pair-wise interactions between restrained particles, hence lower the computational complexity. In order to quantify the computational cost of the force update, we define the force function $\psi: \R \rightarrow \R$ such that $\psi:=v^{\prime} $. Then the \textit{computational cost of the~force update} is defined as the number of times the force function $\psi$ is called. The speed-up of AR dynamics, due to the decreasing of the computational complexity in the force update, with respect to the non-adaptive method which updates all interactions, defines \textit{the algorithmic speed-up}. Since the computational complexity depends on the ratio of restrained particles, which is a quantity that varies at each time step, we consider averages over the whole simulation. More precisely, we denote by $C_{{\rm AR},n}$ the computational cost of the force update in the AR-method at time step $n$ and by $C_{{\rm std},n}$ the computational cost of a standard, non-adaptive method. We denote by $N_{{\rm iter}}=T/\de t$ the number of time steps in the simulation.
Then \textit{the algorithmic speed-up} $S_{\rm a}$ is the ratio of the average computational cost $\widehat{C}_{{\rm std}}:=\E_{\mu_{{\rm std}}} \left[C_{{\rm std}}\right]$ in the standard method and the average computational cost $\widehat{C}_{{\rm AR}}:=\E_{\mu_{{\rm AR}}} \left[C_{{\rm AR}}\right]$ in the AR method:
\begin{equation}
\ds S_{{\rm a}}:=\frac{ \ds {\rm lim}_{N_{{\rm iter}} \rightarrow \infty} \frac{1}{N_{{\rm iter}}} \sum_{n=0}^{N_{{\rm iter}}} {C}_{{\rm std}, n}  }{ \ds{\rm lim}_{N_{{\rm iter}}\rightarrow \infty} \frac{1}{N_{{\rm iter}}}\sum_{n=0}^{N_{{\rm iter}}} {C}_{{\rm AR},n}}=\frac{\widehat{C}_{{\rm std}}}{\widehat{C}_{{\rm AR}}}\,.
\label{algo speed-up definition}
\end{equation}

Note that the computational complexities in both cases are bounded functions of the number of particles and, due to the ergodicity of the dynamics, which was proved by \cite{trstanova2015errorAnalysisOfMLD}, the averages in \eqref{algo speed-up definition} almost surely converge.

However, the important point is the reduction of the error for a given wall-clock duration. We focus here on the statistical error, which is often the dominant source of errors. In order to express the total speed-up with respect to the standard method, we need to consider not only the algorithmic speed-up, but also the modification of the asymptotic variance which depends on the concrete choice of the kinetic energy (see expression \eqref{eq: discrete variance}). We define \textit{the total speed-up} $S_{\rm total}$ as a ratio of the wall-clock time, which is needed by using the AR-method in order to achieve some statistical precision, and the wall-clock time needed for reaching the same precision by the standard method:
\begin{equation}
S_{\rm total}:=\frac{T_{{\rm std}}^{{\rm wlck}}}{T_{{\rm AR}}^{{\rm wlck}}}\,.
\label{eq: def total speed up}
\end{equation}
Recall that, for an observable $\vp$, we denoted by $\sigma^2_{\de t}$ the asymptotic variance of the sampling from the discretized dynamics and by $\sigma^2$ the asymptotic variance of the continuous dynamics. From the Central Limit Theorem, the statistical error at time $T$ is given by
\[
\hat{\vp}_T=\E_{\mu}\left(\vp\right)+ \e_T\mathcal{G},
\]
where $\e_T$ is of order $\frac{\sigma}{\sqrt{T}}$ and $\mathcal{G} \sim \mathcal{N}\left(0,1\right)$. Hence the number of time steps $N_{{\rm iter}}=T/\de t$ needed in order to have a statistical error of order $\e_T$ is
\[
N_{\mr{iter}}=\frac{\sigma^2_{\de t}}{\e_T^2}\,.
\]
The corresponding wall-clock time is therefore obtained by considering the average cost $\widehat{C}$ as
\[
T^{\mr{wlck}}=N_{\mr{iter}} \widehat{C}\,.
\]
Taking into account that $\de t\sigma^2_{\de t}\sim \sigma^2$ (for time steps small enough, recall Section \ref{Computation of macroscopic properties using MD simulations}), the total speed-up $S_{\rm total}$ defined in \eqref{eq: def total speed up} can be expressed as
\begin{equation}
S_{\rm total}=	 \frac{\widehat{C}_{\mr{std}}}{\widehat{C}_{\mr{AR}}}\frac{\sigma^2_{\mr{std}, \de t}}{\sigma_{\mr{AR},\de t}^2}=S_{{\rm a}}\frac{\sigma^2_{\mr{std}, \de t}}{\sigma_{\mr{AR}, \de t}^2}\approx S_{{\rm a}}\frac{\sigma^2_{\mr{std}}}{\sigma_{\mr{AR}}^2}\frac{\de t_{\rm{AR}}}{\de t_{\rm{std}}}\,.
\label{total speed-up}
\end{equation}

The last two terms in \eqref{total speed-up} become equal for small values of $\de t$ and it is therefore sufficient to study the variance of the continuous process $\sigma^2_{\rm{std}}$ and $\sigma^2_{\rm{AR}}$. As we have already mentioned, the choice of the modified kinetic energy should not change the stability properties of the standard dynamics. This would otherwise require us to choose a smaller time step size $\de t_{\rm{AR}}$, which would lead to a smaller total speed-up $S_{\rm total}$. Unfortunately, this is the case of the kinetic energy defined in \cite{PRL-ARPS}. Still, the stability can be significantly improved by using the kinetic energy given by \eqref{def new arps kinetic energy} instead. In this case, the stability properties become comparable to the ones of the standard dynamics (\cite{NumericsInPreparation}). We therefore assume in this work $\de t_{\rm{std}}=\de t_{\rm{AR}}$.

 The computation in \eqref{total speed-up} shows the trade-off between the algorithmic speed-up and the change in variance. Both the algorithmic speed-up $S_{{\rm a}}$ and the AR variance $\sigma^2_{\mr{AR}}$ depend on the parameters of the AR dynamics. As already showed in \cite{PRL-ARPS}, in some applications, the restraining parameters can be chosen such that the total speed-up satisfies $S_{\rm total}>1$. Therefore, there are systems for which this method can be efficient, even though this might be counter-intuitive since one could suggest that in order to accelerate the sampling, the system should move ``faster'' and not be restrained. Note however that the wall-clock duration of the force update step depends on the implementation and on the complexity of the evaluation of $\psi$. Hence, the same physical model with variance $\sigma^2$, can have various algorithmic speed-ups $S_{\rm a}$. Finally, an interesting observation is, that due to the separability of the Hamiltonian, the algorithmic speed-up does not depend on the potential.


\section{Algorithmic speed-up}
\label{section Algorithmic speed-up}

The goal of this section is to propose a methodology to analyze the algorithmic speed-up $S_{\rm a}$ (defined in~\eqref{algo speed-up definition}) of AR dynamics as a function of the percentage of restrained particles. We first describe the adaptive algorithm for computing forces, and we estimate the corresponding computational cost. In the last part, we also consider the effort for updating neighbor lists used for updating of short-ranged interactions and we obtain an estimation of the algorithmic speed-up per time step.

\subsection{Description of the AR force update algorithm}

For simplicity, we consider a system of $N$ particles where only pair-wise interactions take place. In AR dynamics, this sum can be split into three kinds of interactions depending on the state of the two interacting particles: active-active, active-restrained and restrained-restrained. We define the set of indices of active particles $I_{\mr{A}}$ and restrained particles $I_{\mr{R}}$. Then, sum \eqref{force on one particle} can be re-written as
\begin{equation}
f_i=\sum_{\substack{j\in I_{\rm{A}} \\ j\neq i}}f_{ij}+\sum_{\substack{j\in I_{\rm{R}} \\ j\neq i}}f_{ij}\,.
\label{eq: force i sum active restrained}
\end{equation}
The force acting on particle $i$ in the next time step $n+1$ can be formally obtained using the old position $q^n$:
\begin{equation}
f_i^{\rm{new}}=f_i^{\rm{old}}+ \left(f_i^{\rm{new}}-f_i^{\rm{old}}\right)\,, \quad f_i^{\rm{old}}=\sum_{j}f_{ij}(q^n), \quad f_i^{\rm{new}}=\sum_{j}f_{ij}(q^{n+1})\,.
\label{eq: force i sum active restrained 2}
\end{equation}
Since, for the set of restrained particles, positions have not changed since the previous time step, one can easily see that
\[
\forall i\in I_R,\quad\sum_{\substack{j\in I_{\rm{R}} \\ j\neq i}}f_{ij}^{\rm{new}}-f_{ij}^{\rm{old}}=0\,.
\]
The computation in \eqref{eq: force i sum active restrained 2} is therefore reduced to subtracting the old and adding the new active-restrained and active-active interactions. This simple remark provides in fact a key point for the reduced complexity of the AR algorithm.

In a standard simulation, when taking into account Newton's third law $f_{ij}=-f_{ji}$, the computational cost of pair-wise interactions is $\frac{N(N-1)}{2}$. The resulting quadratic complexity in the number of particles is not favorable due to the system size, and therefore neighbor lists are usually introduced (see \cite{allen1989computer,frenkel2001understanding}). For a comparison of various approaches for neighbor list methods we refer the reader to \cite{artemova2011comparison}. Neighbor lists can be used in systems where forces vanish after a certain cut-off distance, so that each particle only interacts with a relatively limited number of neighbors. For simplicity, we consider a homogenous system where we assume that the number of neighbors $C$ of a particle is the same for each particle. Taking into account that, for each pair $(i, j)$, we may only compute the force $f_{ij}$ and deduce $f_{ji}$ thanks to Newton's third law, the number of interactions reduces to $\frac{NC}{2}$.

The BAO discretization scheme \eqref{discr modified Langevin first order} can be formalized in the following way:
\begin{mdframed}
\begin{algorithm}[H]
 \textbf{Input}: {Initial conditions $p^0, q^0$}\\
  \textbf{Output}: {$p^n, q^n$}\\
  \For{each time step and each particle}{
  B: Update momenta \;
	A: Update positions \;
	\quad \ Update neighbor-lists \;
		\quad \ Update forces \;
	O: Update momenta in fluctuation-dissipation part (FD) \;
   }
 \caption{Algorithm for \eqref{discr modified Langevin first order} in the case of the standard dynamics.}
\label{algo std 1}
\end{algorithm}
\end{mdframed}

In AR dynamics, the information about which particles are going to move after the position update is already available after updating the momenta B, since the kinetic energy will not change anymore. The algorithm above may thus be modified in the following way:

\begin{mdframed}
\begin{algorithm}[H]
 \caption{Algorithm for \eqref{discr modified Langevin first order} in AR dynamics using adaptive forces updates.}
\textbf{Input}: {Initial conditions $p^0, q^0$}\\
\textbf{Output}: {$p^n, q^n$}\\
	For each particle $i$, initialize force $f_i$; \\
 \For{each time step and each particle}{
 	 B: Update momenta; \\
		\quad \ Create lists of active and restrained particles; \\
		\quad \  Subtract active-active and active-restrained interactions; \\
	 A: Update positions; \\
	\quad \ Update neighbor-lists; \\
		 \quad \ Add active-active and active-restrained interactions; \\
		O: Update momenta in FD;
 }
\label{algo arps 1}
\end{algorithm}
\end{mdframed}

Updating neighbor lists normally consists in re-assigning each particle to a specific grid cell (in our implementation we used a combination of Verlet lists and linked-cell lists (\cite{frenkel2001understanding})). In AR dynamics, restrained particles do not have to be re-assigned, and neighbor lists may be updated more efficiently. More precisely, the complexity of updating the neighbor lists goes from $O(N)$ the number of particles, to $O(K)$, where $K$ is the number of active particles.

Note that the force function $\psi$ is called in both AR force updates (subtract and add steps), since we need to evaluate forces for positions at the previous time step. It would be possible to avoid updating forces twice by saving all pairwise forces, but this may result in a quadratic space complexity. We will not analyze this case, although it would result in a larger algorithmic speed-up and lead to less restrictive conditions on the efficiency of AR dynamics.

Note that there is a slight overhead due to computing the AR kinetic energy functions $\n K$, which is more complicated than in the standard case. Still, this involves $O(N)$ additional operations, and can be neglected compared to the cost of the force update in practical applications. Furthermore, the overhead mostly comes from the transition regime since $\n K$ vanishes for restrained particles and becomes $M^{-1}p$ for the full-dynamics state.

Note that a similar strategy for incremental force update may be applied to other splitting schemes of the modified Langevin equations. However, the status of a particle (active, in transition or restrained) depends on the state of the momenta before the position update, and hence this status should not be destroyed by updating momenta between two position updates. Using the same notation as in Section~\ref{Computation of macroscopic properties using MD simulations} and \cite{Matthews}, this implies that the second order splittings BAOAB and OABAO are not directly suited for a modification by the AR dynamics algorithm, since between the two position updates A, the momentum changes by either O or B step. On the other hand, ABOBA, BOAOB, OBABO and AOBOA can be used, since the lists of active particles can be created before the position update A and hence active-active and active-restrained interactions can be subtracted and added after position update A.

\subsection{Complexity analysis}

At each time step, the computational cost of the force update depends on the percentage of restrained particles. Let us denote the number of active particles by $K=\alpha N$, where $\alpha\in \left[0, 1\right]$ is the average ratio of active particles. The number of restrained particles is then $N-K$. We are going to formalize the computational complexity of the force update as a function of the ratio of restrained particles, denoted by $\rho:=1-\alpha $.

We recall that we have considered the average computational cost over the whole trajectory in equation \eqref{algo speed-up definition}, since the instantaneous computational cost may vary at each time step. Because, in the algorithm analyzed in this paper, we add and subtract pairwise forces, the computational complexity of the force update in an AR simulation is lower than a regular force update if and only if a sufficient number of particles is restrained. We are thus going to analyze which conditions on the number of restrained particles are sufficient to obtain a speed-up larger than one, when a standard simulation has a linear or quadratic complexity\footnote{The quadratic complexity corresponds to bonded interactions and the linear complexity to non-bonded, in which case the neighbor lists can be applied. }. This analysis can be extended to other force update algorithms.

\subsubsection{Quadratic complexity}

Let us first consider a standard (non-adaptive) simulation with a quadratic-complexity force update algorithm, i.e. when no neighbor-lists are used. In this case, the number of interactions computed at every time step is $C_{\rm std}:=\frac{N(N-1)}{2}$. In AR dynamics, we do not need to recompute interactions between restrained particles, hence we only update interactions involving active particles, either with other active particles, or with restrained particles. As a result, the computational cost for the AR force update is\footnote{The factor of 2 comes from the need to subtract old forces (with previous positions) and add new forces (with current positions).}:
\[
C_{\rm AR}:=2\left(\frac{K(K-1)}{2}+K(N-K)\right)=\left(2N-\alpha N-1\right)N\alpha\,.
\]
and $C_{\rm std}>C_{\rm AR}$ is satisfied for
\begin{equation}
\label{necessary condition no neighbor lists}
\alpha < 0.29
\end{equation}
 and $N>\frac{1-2\alpha}{2\alpha^2-4\alpha+1}$. The inferior bound on the number of particles is not a restrictive condition for molecular dynamics, where the number of particles is usually much bigger. (For example, for $\alpha=0.28$, the number of particles $N$ must be larger than $12$.)  More importantly, this implies that at least $71\%$ of particles must be restrained in order for this force update algorithm to be beneficial, in which case the algorithmic speed-up is:
\[
S_{{\rm a},1}= \frac{C_{\rm std}}{C_{\rm AR}}= \frac{N-1}{2(N\rho+N-1)(1-\rho)}\,.
\]
When the number of particles tends to infinity, this becomes
\begin{equation}
S_{{\rm a},1}^{\infty}=\lim_{N \rightarrow \infty}S_{{\rm a}}(N)=\frac{1}{2(\rho+1)(1-\rho)}\,.
\label{eq: algo speed-up no neighbor lists}
\end{equation}
 Note that if the double computation of forces can be avoided (for example by storing previous pairwise forces), the complexity becomes
\[
C_{{\rm AR},2}:=\frac{C_{\rm AR}}{2}\,,
\]
so that $C_{{\rm AR},2}>C_{\rm std}$ is achieved for any $\alpha <1$ and $N>\frac{1}{1-\alpha}$, resulting in the following speed-up:
\[
S_{{\rm a},2}=\frac{N-1}{\left(N\rho+N-1\right)\left(1-\rho\right)}, \quad S_{{\rm a},2}^{\infty}=\frac{1}{(\rho+1)(1-\rho)}\,.
\]

\subsubsection{Linear complexity}

Let us now consider the (much more frequent) case where the complexity of the force update is linear, e.g. when forces become sufficiently small after a given cutoff distance $r_{\rm cut}$, and neighbor lists may be used to efficiently determine which particles are interacting. The reference complexity is therefore $C_{\rm{std, NL}}=\frac{NC}{2}$, where $C$ is the (average) number of neighbors. The algorithm for the adaptive force update is as follows: for all active particles compute interactions with their neighbors, and between the active neighbors use $f_{ij}=-f_{ji}$. The total number of interactions to be updated in the AR dynamics algorithm is then:
\begin{equation}
C_{{\rm AR},\mr{NL}}:=2\left(\sum_{i=1}^K\sum_{j\in L_A(i)} 1 + \sum_{i=1}^K\sum_{j\in L_R(i)} 1 \right)=2\left(\frac{KC_A}{2}+KC_R\right)=\left(1-\frac{\alpha}{2}\right)\alpha CN \,,
\label{cost NL}
\end{equation}
where the set $L_A(i) \subset I_A$ contains the indices of the active neighbors of the particle $i$, $L_R(i)\subset I_R$ contains the indices of the restrained neighbors, $C_A=\alpha C$ and $C_R=\rho C$.
 The necessary condition for $C_{\rm{std, NL}}>C_{\rm AR,{NL}}$ is then
\begin{equation}
\label{eq: condition arps nl}
\alpha < 0.293\,. 
\end{equation}
Note that this condition does not depend on $N$, nor on  $C$. The AR dynamics algorithm is more efficient in number of operations for forces update if and only if the percentage of restrained particles is bigger than $70,7\%$. The algorithmic speed-up 
 is hence
\begin{equation}
S_{\rm{NL}} := \frac{C_{\rm std}}{C_{\rm AR, NL}}=\frac{1}{2\alpha(2-\alpha)}=\frac{1}{2(1-\rho^2)}\,.
\label{eq: algorithmic speed-up}
\end{equation}
Again, avoiding the double re-computation of force components from the old positions for the active particles, removes a factor of 2 from $C_{\rm AR, NL}$ and the computational cost becomes $C_{\rm AR, NL}=\left( \frac{1}{2} -\alpha\right)\alpha CN$, which implies an unconditional algorithmic speed-up $S_{\rm a}=\left(1-\rho^2\right)^{-1}$.

An important conclusion is that an incremental force update is computationally beneficial if the percentage of restrained particles is larger than a threshold $\mathcal{R}$. We may thus modify Algorithm \ref{algo arps 1} as follows:
		
\begin{mdframed}
	\begin{algorithm}[H]
 \textbf{Input}: {Initial conditions $p^0, q^0$}.  \textbf{Output}: {$p^n, q^n$}\\
For each particle $i$, initialize the force $f_i$; \\
 \For{each time step and each particle}{
 		 B: Update momenta; \\
		 \quad \ Create lists of active and restrained particles; \\
		 \eIf {$\rho > \mathcal{R}$ }{
		 \quad \ Subtract active-active and active-restrained interactions; \\
		A: Update positions; \\
		 \quad \ Add active-active and active-restrained interactions; \\
		}
		{		
		A: Update positions; \\
		\quad \ Update forces with the standard approach; \\
		}
		\quad \ Update neighbor-lists; \\
		O: Update momenta in FD; \\
 }
\caption{Improvement of Algorithm \ref{algo arps 1} by using the condition on the ratio of restrained particles $\rho$ given by the constant $\mathcal{R}\geq 0$.}
\label{algo arps 2}
\end{algorithm}
\end{mdframed}

Finally, we consider the case where the neighbor lists are updated at each time step\footnote{Note that this can be easily modified in order to express the update of neighbor-lists every time period $T$.}. This is not usually done in practical applications, where neighbor lists are updated only after a certain time period which can be computed from the maximal velocity of the particles. The cost per time step then extends in re-assigning $N$ particles into the grid, which gives order of $NC/2 + N$ operations. In the AR simulation, only active particles need to be re-assigned into the grid. Therefore, the cost per time step computed in \eqref{cost NL} becomes
\[
C_{\rm AR,\mr{NL}}:=\left(1-\frac{\alpha}{2}\right)\alpha CN +\alpha N\,.
\]
	Assuming that there are $C$ neighbors in average, the resulting speed-up is:
	\begin{equation}
	S_{\rm a}=\frac{C+2}{2(1-\rho)(C\rho+C+1)} \,.
	\label{eq: speed up newton nl}
	\end{equation}

\section{Total speed-up}
\label{section Total speed-up}

As explained above, the total speed-up \eqref{algo speed-up definition} reachable by AR dynamics when estimating observables depends on both the computational complexity of the force update, and the variance of the AR dynamics.

In this section, we first analyze how the percentage of restrained particles depends on the restraining parameters $\er$ and $\ef$.
Then, we approximate the variance of the AR dynamics by a linear function. Combining both, we finally express the total speed-up as a function of $\er$ and $\ef$.

\subsection{Percentage of restrained particles}
The percentage of restrained particles can be computed from the average occupation of the restrained state of each particle. In other words, it is the probability that the momenta of one particle belong to the restrained region of phase space.
For the AR kinetic energy function \eqref{def new arps kinetic energy}, the average occupation of the restrained state $R(\er,\ef)$ of particle $i$ with parameters $\er$ and $\ef$ is the expected value of the indicator function of the absolute values of all momenta components of one particle being smaller than the restraining parameter $\er$.

We denote by $\mu_{\er,\ef}$ the invariant measure which corresponds to the AR kinetic energy function with parameters $\er$ and $\ef$ and we compute
\begin{equation}
R(\er,\ef)=\int_{\R^d}\1_{\left\{\frac{\left|p_i\right|}{m_i}\leq \er\right\}}\mu_{\er,\ef}=\frac{\ds \left(2\er m_i\right)^D{\rm{exp}}\left({-\beta D S_{\er,\ef}}\right)}{Z_p(\er,\ef)}\,,
\label{eq: RerefNew}
\end{equation}
where the momenta normalization constant of the particle $i$ is 
simply $Z_p=z^D$, with
\begin{equation}
\begin{aligned}
z(\er,\ef)&=\int_{\left\{\frac{\left|p_i\right|}{m_i}\leq \er\right\}}{\rm e}^{-\beta S_{\er,\ef}} \ dp+\int_{\left\{\frac{\left|p_i\right|}{m_i}\geq \ef\right\}}{\rm e}^{-\beta \frac{p_i^2}{2 m_i}} \ dp\\
&\qquad + \int_{\left\{\frac{\left|p_i\right|}{m_i}\in \left[\er,\ef\right]\right\}}{\rm e}^{-\beta s(p_i)}dp\\
&=\left(2\er m_i\right){\rm e}^{{-\beta S_{\er,\ef}}}+\sqrt{2\pi m_i\beta^{-1} } - 2 \int_{0}^{ m_i  \ef} {\rm{e}}^{-\beta\frac{p^2}{2m_i}} dp \\
&\qquad + \int_{{m_i\er}}^{{m_i\ef}} {\rm{e}}^{-\beta s\left(p_i\right)} dp+ \int_{{-m_i\ef}}^{{-m_i\er}} {\rm{e}}^{-\beta s\left(p_i\right)} dp\,.
\end{aligned}
\label{Z_p er ef}
\end{equation}
Note that in the standard dynamics $Z_p=(2\pi m_i/\beta )^{3/2}$.

Finally, considering a system consisting of particles with various restraining parameters $\er^i$ and $\ef^i$, the total average percentage of restrained particles can be computed as an average over the individual values $R({\er^i,\ef^i})$ of each particle. Denoting by $N_{\er,\ef}$ the number of particles with parameters $\er$ and $\ef$ and by $\mathcal{N}$ the set of all chosen pairs $(\er,\ef)$, the total average percentage of restrained particles\footnote{Note that this corresponds to the notation $\rho=R_{\rm{total}}$ in Section \ref{section Algorithmic speed-up}.} $R_{\rm{total}}\in \left[0,1\right]$ is given by
\begin{equation}
R_{\rm{total}}= \frac{1}{N}\sum_{\left(\er,\ef\right)\in \mathcal{N}}N_{\er,\ef}R(\er,\ef) \,.
\label{total speed-up as perc restr part}
\end{equation}
For example, the percentage of restrained particles for a system consisting of a dimer that follows standard dynamics and that is surrounded by solvent particles following AR dynamics with non-zero parameters $\er$ and $\ef$ is:
\[
R_{\rm{total}}^{\rm{DS}}= \frac{N_{\rm{Solv}}}{N_{\rm{total}}}R_{\rm{Solv}}(\er,\ef)\,,
\]
since, in standard dynamics, the average occupation of the restrained state is zero and $R_{\rm{Dimer}}=0$.

In conclusion, the algorithmic speed-up $S_{\rm a}$ can be estimated using the computational complexity of the algorithm (see Section \ref{section Algorithmic speed-up}) with the speed-up being a function of $\rho=R_{\rm total}$.

\begin{figure}[h]
	\centering
		\includegraphics[width=0.80\textwidth]{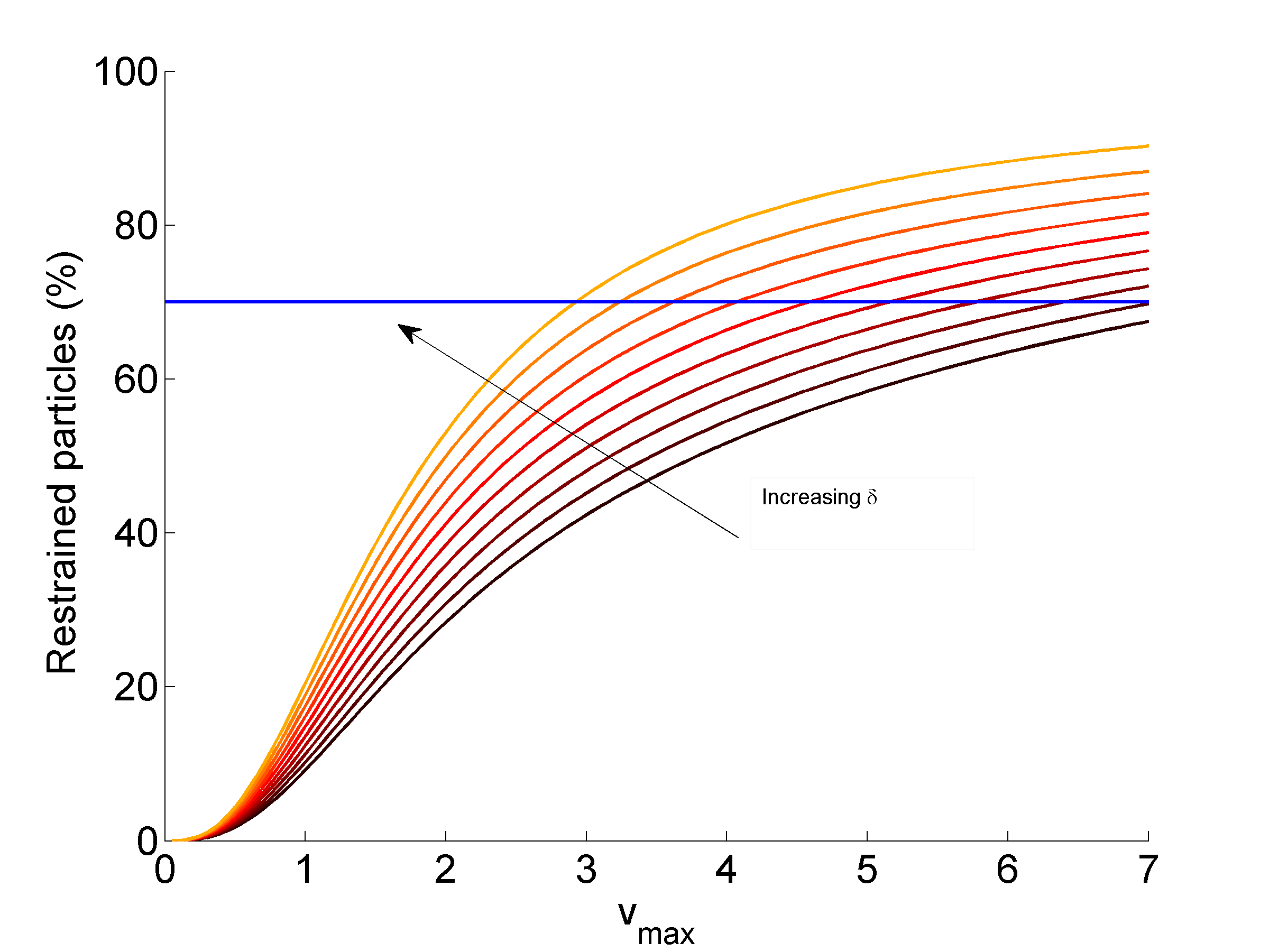}
	\caption{\textit{Average occupation of the restrained state with respect to the parameter ratio.} We computed $R_{\rm{total}}=R(\er,\ef)$  for one particle in 3D according to \eqref{eq: RerefNew} for various $\ef$ and various values of the parameter ratio $\delta\in\left[0.7,0.98\right]$  (black to orange or bottom to top lines) such that $\er=\delta\ef$. The blue line is the value $70\%$ of restrained particles which corresponds to the necessary condition for $S_{\rm a}>1$.}
	\label{fig:percRestrOverKmaxForVariousDelta3d}
	\end{figure}

\begin{figure}[h]
	\centering
			\includegraphics[width=0.80\textwidth]{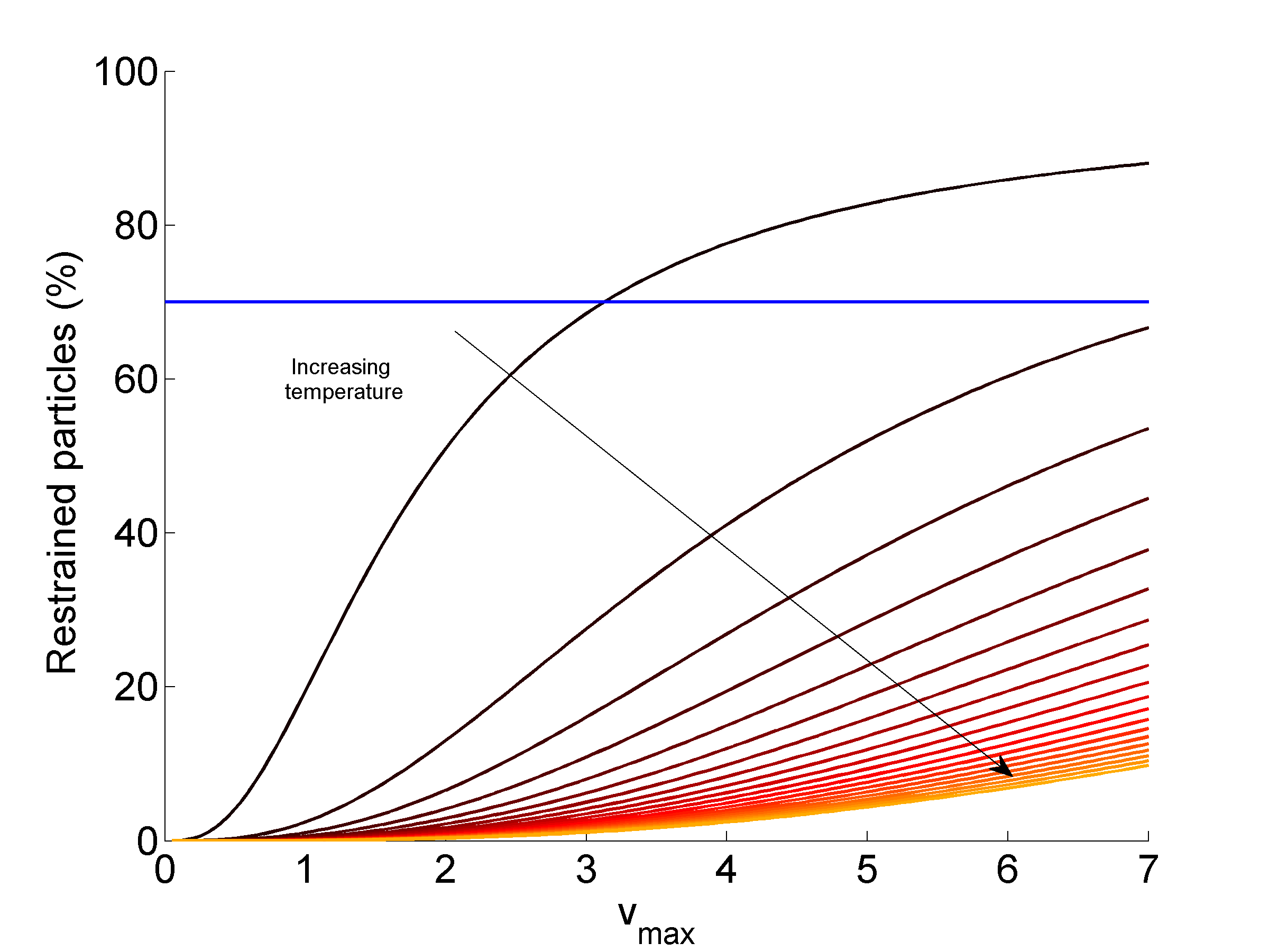}
			\caption{\textit{Average occupation of the restrained state with respect to the temperature.} We computed $R_{\rm{total}}=R(\er,\ef)$ of a particle in 3D according to \eqref{eq: RerefNew} for $\ef$ and $\er=0.95\ef$ and for various temperatures $k_{B}T=[1, 100]$ (black to orange or top to bottom lines).}
	\label{fig:percentage_restr_part_wrt_temperature}
	\end{figure}

Figure \ref{fig:percRestrOverKmaxForVariousDelta3d} shows, for $K$ defined in \eqref{def new arps kinetic energy}, the average occupation of the restrained state $R(\er,\ef)$ as a function of $\ef$ for various $\delta\in \left[0.5,0.98\right]$ such that $\ef=\delta \er$ in dimension three. We depicted also the value $70\%$ of restrained particles which corresponds to the necessary condition for $S_{\rm a}>1$ (given by \eqref{necessary condition no neighbor lists} or \eqref{eq: condition arps nl}). We observe on this figure that the bigger $\delta$, the bigger average occupation of the restrained state. Figure \ref{fig:percentage_restr_part_wrt_temperature} shows the dependence of $R(\er,\ef)$ on the temperature. This suggests that the restraining parameters should be scaled with respect to the temperature $k_BT$.

\begin{figure}[h]
	\centering
			\includegraphics[width=0.80\textwidth]{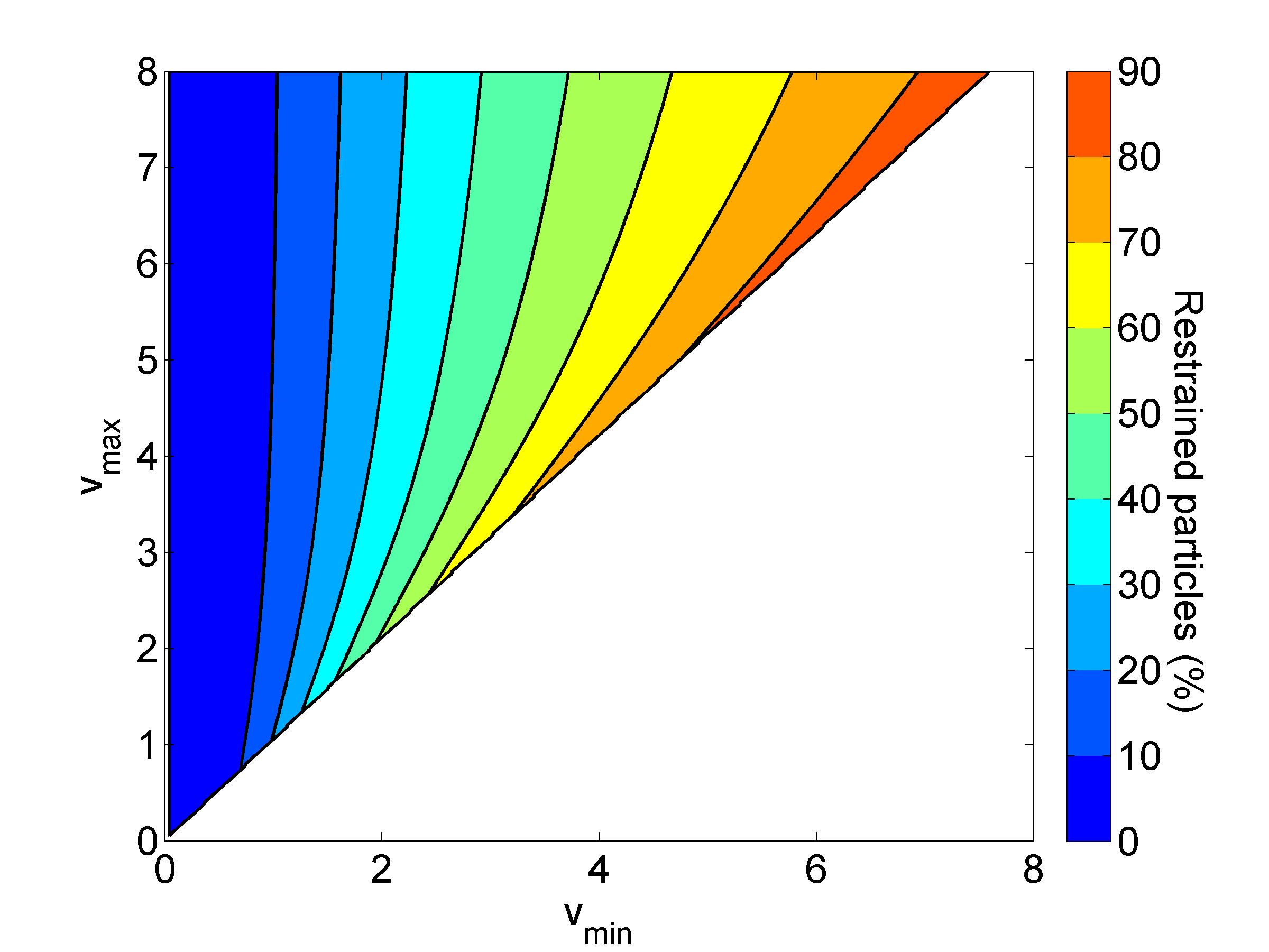}
			\caption{\textit{Percentage of restrained particles over $\er$ and $\ef$.} We considered the system of a dimer surrounded by a solvent (64 particles) in 3D  with the solvent dynamics being governed by the AR kinetic energy function. We computed the percentage of restrained particles for various values $0\leq\er\leq 0.95\ef$. }
			\label{fig:percRestrPart}
		\end{figure}

Finally, Figure \ref{fig:percRestrPart} shows $R_{\rm{total}}(\er,\ef)$ as a function of both parameters. Note that the highest value of percentage of restrained particles is located close to the diagonal, \textit{i.e.} when the gap between the parameters $\er$ and $\ef$ is small.

\subsection{Linear approximation of the variance}

In \cite{trstanova2015errorAnalysisOfMLD}, it was proved that 
there exists a regime in which the variance from the AR dynamics simulations can be approximated by a linear function of the restraining parameters (see \cite[Proposition 4.3]{trstanova2015errorAnalysisOfMLD}):
there exists $\ef^*$ small enough such that for $\ef<\ef^*$ there exist constants $c_1,c_2\in \R$ such that for $\er\in\left[0,\ef\right]$
\begin{equation}
\sigma^2_{\mr{AR}}(\er+\zeta,\ef+\eta)=\sigma^2_{\mr{AR}}(\er,\ef)+c_1\zeta +c_2\eta+\mr{O}\left(\zeta^2+\eta^2\right).
\label{eq: variance linear behavior}
\end{equation}

The total speed-up of the wall-clock time needed to reach a certain statistical precision \eqref{total speed-up} can hence be expressed in terms of the restraining parameters using \eqref{eq: variance linear behavior} as 
\begin{equation}
\begin{aligned}
S_{\rm total}&\approx S_{{\rm a}}(\er,\ef)\frac{\sigma^2(0,0)}{\sigma^2(0,0)+c_1\er +c_2\ef}\\
&\quad=S_{{\rm a}}(\er,\ef)\frac{1}{1+\frac{c_1}{\sigma^2(0,0)}\er +\frac{c_2}{\sigma^2(0,0)}\ef}.
\end{aligned}
\label{eq: total speed-up approx}
\end{equation}
The gap between the restraining parameters $\er$ and $\ef$ should be big enough to ensure a smooth transition between the full and the restrained dynamics and prevent numerical instabilities. Note, however, that in the numerical experiments performed in \cite[Section 5.1]{trstanova2015errorAnalysisOfMLD}, where the variance was computed for a simple 1D system, it was shown that the relative increase of the variance with respect to the full-dynamics parameter $\ef$ is more significant than with respect to the restraining parameter $\er$. This result is not surprising, since the gap between the parameters smooths out the dynamics, which translates into an increase of correlations. This implies that the optimal strategy is to choose the gap between the parameters as small as possible while still maintaining the numerical stability and keeping the systematical error sufficiently low (i.e. the error on the computed averages, arising from the fact that $\mu\neq \mu_{\de t}$ (\cite{Matthews})). At the same time, the restraining parameters should give the desired percentage of restrained particles $R_{\rm{total}}$. For example, in the case when $\delta=\er/\ef=0.98$, 
the relative derivative of the restrained energy of one particle $R(\er,\ef)$ with respect to $\ef$, almost vanishes after the value $\ef=5$. Hence this is a critical value after which the growth of function $R(\er,\ef)$ slows down 
 (see again Figure \ref{fig:percRestrOverKmaxForVariousDelta3d}).	
 Having in mind that the variance locally increases with respect to $\ef$, this implies that, in this region, the efficiency of the algorithmic speed-up does not grow fast enough with increasing $\ef$, while the variance might be still growing. In this case, either the gap $\delta$ should be chosen smaller, or one must ensure that the variance does not grow too fast, in order to compensate the variance growth with the algorithmic speed-up.

It is easy to estimate the algorithmic speed-up $S_{{\rm a}}$. The problematic part is to estimate the sensitivity of the variance of a given observable with respect to the modification by the restrained dynamics, i.e. the estimation of $c_1$ and $c_2$ in \eqref{eq: total speed-up approx}. This can be done by a linear interpolation in the pre-processing part, which should involve at least three AR dynamics simulations in order to estimate the constants $c_1$ and $c_2$. 
Finally, \eqref{eq: total speed-up approx} allows to have a complete expression for the total speed-up as a function of the parameters $\er$ and $\ef$. Choosing $\delta=\er/\ef$ as small as possible, one can find the optimal $\ef$ which produces the highest total speed-up (see Section \ref{section numerical examples} for a numerical example).

We thus propose the following guidelines to estimate the total speed-up $S_{\rm total}$ with respect to the parameters $\er$ and $\ef$:
\begin{mdframed}
\begin{enumerate}
\item Choose the order (scale) of the restraining parameters $\er$ and $\ef$ for each particle according to its mass, its role in the system and the temperature $k_BT$.
\item Choose the minimal gap $\delta$ between $\er$ and $\ef$ with respect to the numerical stability.
	\item Compute the percentage of restrained particles $R_{\rm{total}}$ according to \eqref{eq: RerefNew}, \eqref{Z_p er ef} and \eqref{total speed-up as perc restr part}.
	\item Compute the algorithmic speed-up $S_{{\rm a}}$ according to the implementation algorithm according to Section \ref{section Algorithmic speed-up}.
		\item Estimate the linear approximation of the variance $\sigma^2_{\rm AR}(\delta\ef,\ef)$ for the observable~$A$.
		\item Find the optimal value of $\ef$ (with $\er=\delta \ef$) by maximizing $S_{\rm total}(\er,\ef)$.
\end{enumerate}
\end{mdframed}

\section{Numerical illustration}
\label{section numerical examples}

\begin{figure}[h]
	\centering
			\includegraphics[width=0.80\textwidth]{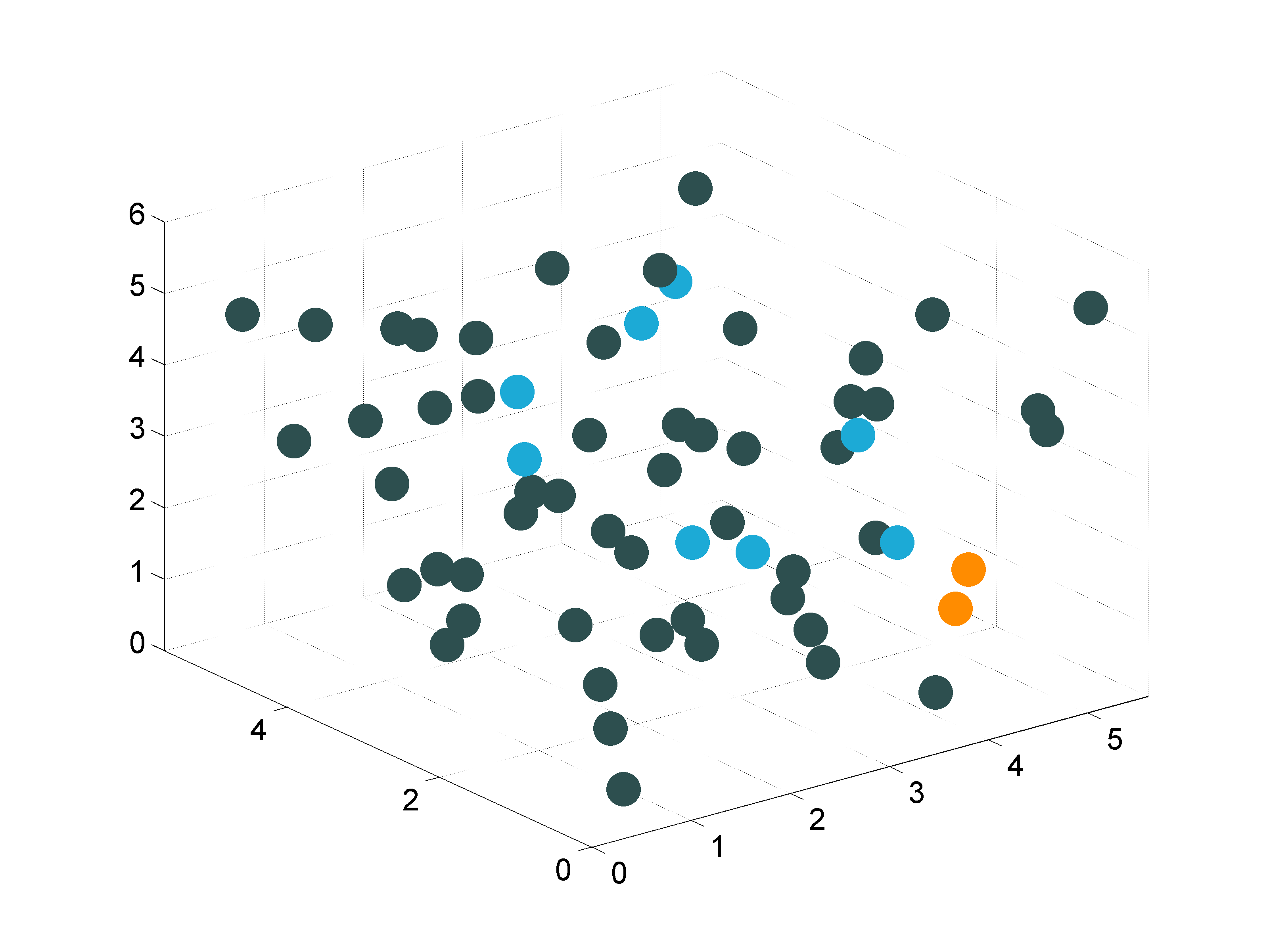}
			\caption{A snapshot from the AR-dynamics simulation of a dimer (orange) surrounded by 62 solvent particles (light blue and grey) with parameters $\er=6.4$ and $\ef=8$ (only for the solvent particles, the dimer has the standard kinetic energy $k_{\rm std}(p_1))+k_{\rm std}(p_2))$). The solvent particles are in two states: restrained (grey) or active (light blue).  }
	\label{fig:3d}
	\end{figure}

In order to illustrate the theoretical results from the previous section, we consider a system of $N=64$ particles consisting of a dimer ($q_1, q_2$) surrounded by  
\nrSolvent solvent particles ($q_3, \ldots, q_N$) in space dimension $D=3$ (see Figure~\ref{fig:3d} for an illustration of this system). This model is representative of many applications in biology, chemistry or physics, where the macroscopic property only depends on a small part of the simulated system. For example, in simulation of a protein in solvent the interest lies in computation macroscopic properties of the protein (see for instance \cite{PRL-ARPS}). The validation of this method for real-world problems is still needed and this work has already started by the ARPS method being implemented into LAMMPS (see \cite{SinghRedon2016}).

We use periodic boundary conditions with box-length 
such that the density is $0.4$. 
 We consider reduced units such that particles have identical masses $m_i=1$ and the temperature is chosen so that $\beta=1$.
The friction constant in the Langevin equations is $\gamma=1$. Solvent particles interact with each other and with the dimer particles by a truncated Lennard-Jones potential with parameter $\e_{\rm{LJ}}=1$ with a cut-off distance $r_{\rm{LJ}}=2^{1/6}$. Dimer particles interact with each-other with a double-well potential (with width $w=1$ and height $h=1$). This potential corresponds to a metastable system, with the two metastable states: compact and stretched.  
For a more precise formulation, see Section 5.2 in \cite{trstanova2015errorAnalysisOfMLD} or \cite{lelievre2010free}. We discretize the modified Langevin equations \eqref{modified Langevin} by a second-order scheme (OBABO) with time step size $\de t=0.001$ and perform $N_{\rm iter}\approx 10^9$ time steps.

We use neighbor-lists based on the cut-off distance of the Lennard-Jones potential $r_{\rm{LJ}}$, according to Algorithm \ref{algo arps 2}. The average number of neighbors is estimated as $C=0.25$. We run one reference simulation in the standard dynamics.

In the AR simulations, we consider non-zero restraining parameters on the solvent only, and we let dimer particles follow the standard dynamics. In order to demonstrate the dependence of the total speed-up $S_{\rm total}$ on the restraining parameters $\er$ and $\ef$, we consider the following observables: the dimer distance $A_D(q_1, q_2)=\left|q_1-q_2\right|$, the dimer potential $A_V(q_1, q_2)=V_{DW}(\left|q_1-q_2\right|)$ and the kinetic temperature $\mathcal{T}(p)=p\cdot \n K(p)$. The first two observables only depend on the positions of the dimer particles, hence we expect that the variance will not be much modified even for large restraining parameters. The function $\mathcal{T}(p)$ depends on the momenta of all particles $p$ and satisfies~$\left\langle \mathcal{T}\right\rangle_{\mu_{\er,\ef}}=k_{B}T$. The knowledge of the exact average allows us to verify that the time step size $\de t$ is chosen sufficiently small in order to make the systematic error on the averages smaller than $1\%$ even for $\ef=10$. The asymptotic variance $\sigma^2$ of a time average for a given observable $\vp$ is estimated by approximating the integrated auto-correlation function by a trapezoidal rule (see Section 5.2 in \cite{trstanova2015errorAnalysisOfMLD}).

\begin{figure}[h]
	\centering
			\includegraphics[width=0.80\textwidth]{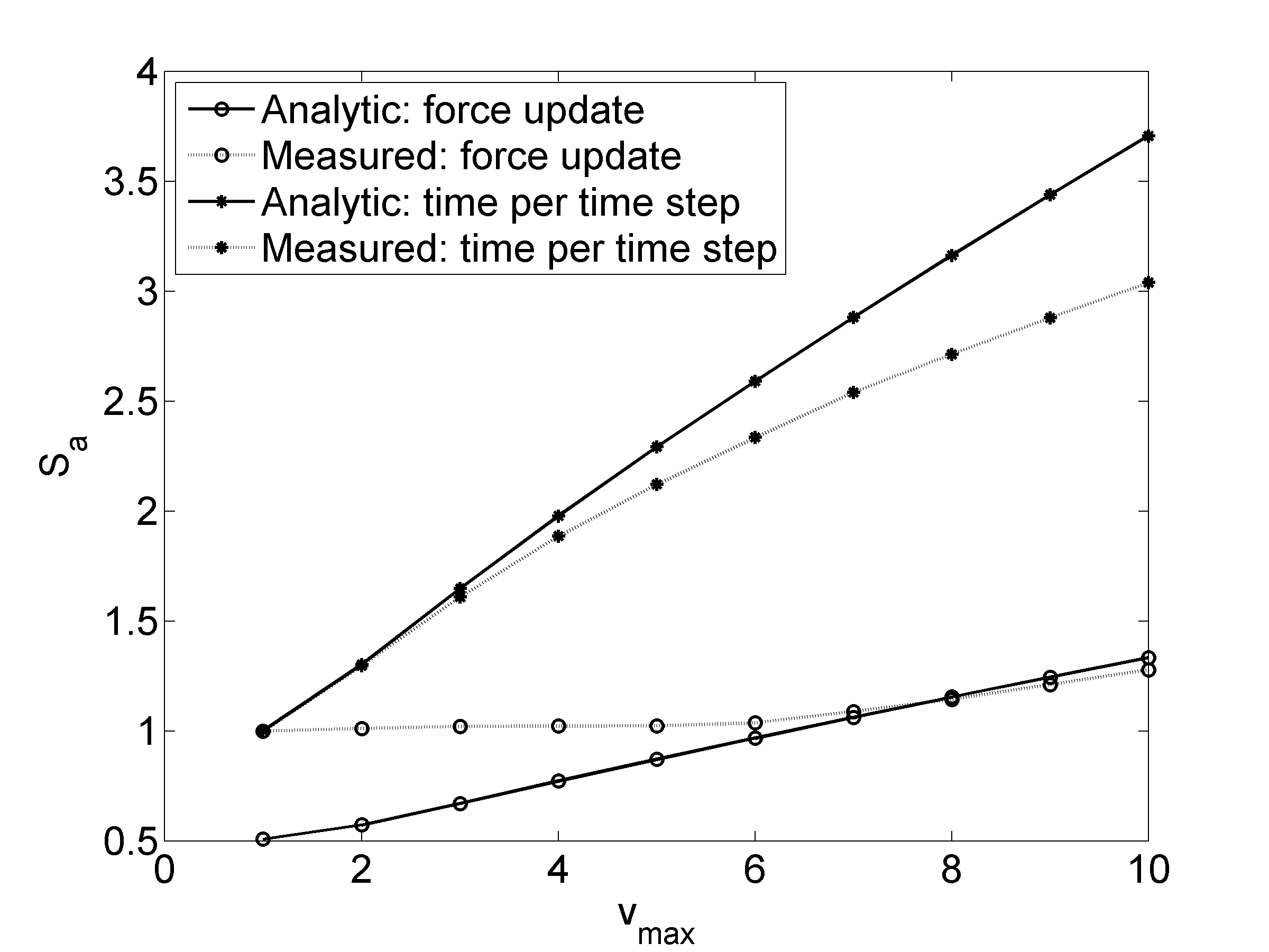}
			\caption{\textit{Comparison of the analytically estimated algorithmic speed-up with the measured one.} We considered a system described in Section \ref{section numerical examples}. We measured the algorithmic speed-up $S_{\rm a}$ of the forces update function and the total time step, with respect to the parameters $\ef$ and $\er=0.8\ef$. We observe the same behavior of the computed speed up in forces update (the one predicted by \eqref{eq: algorithmic speed-up}) and the measured one, as well as the algorithmic speed-up per time step \eqref{eq: speed up newton nl}. Note that the measured $S_{\rm a}$ of the forces update is equal to one for small values of $\ef$, which is due to the implementation of the condition on the adaptive forces update as proposed in Algorithm~\ref{algo arps 2} which assures $S_{\rm a}\geq 1$.}
	\label{fig:comparisonSpeedUpTimePerTimeStepAndForceUpdateNice}
	\end{figure}

First, we confirm theoretical predictions for the algorithmic speed-up $S_{\rm a}$. In our simulations, we measure the time per force update, as well as the time per time step. We compare the measured speed-up, which is a ratio of the measured time in the standard dynamics and the AR dynamics, with the estimated speed-up in the force update \eqref{eq: algorithmic speed-up} and for the overall time step \eqref{eq: speed up newton nl}. Figure~\ref{fig:comparisonSpeedUpTimePerTimeStepAndForceUpdateNice} shows a comparison of the predicted algorithmic speed-up and the measured algorithmic speed-up in our simulation, which demonstrates that it is possible to roughly estimate the computational behavior of a specific implementation. Even for our simple implementation, however, the mismatch in the curves may have multiple causes: the update of positions and momenta, the creating of lists of active particles, the updating of the neighbor-lists \textit{etc.}, and the fact that with growing $\ef$ and a fixed $\delta$, the particle spends more time in the transition region which is computationally more expensive due to the spline. As we suggest later in the paper, realistic implementations such as those found in popular MD packages (e.g. LAMMPS, GROMACS, etc., see Singh and Redon 2016) are very complex, and the best strategy to determine AR parameters may be to actually measure algorithmic speed-ups on short simulations.
%

\begin{figure}[h]
	\centering
			\includegraphics[width=0.80\textwidth]{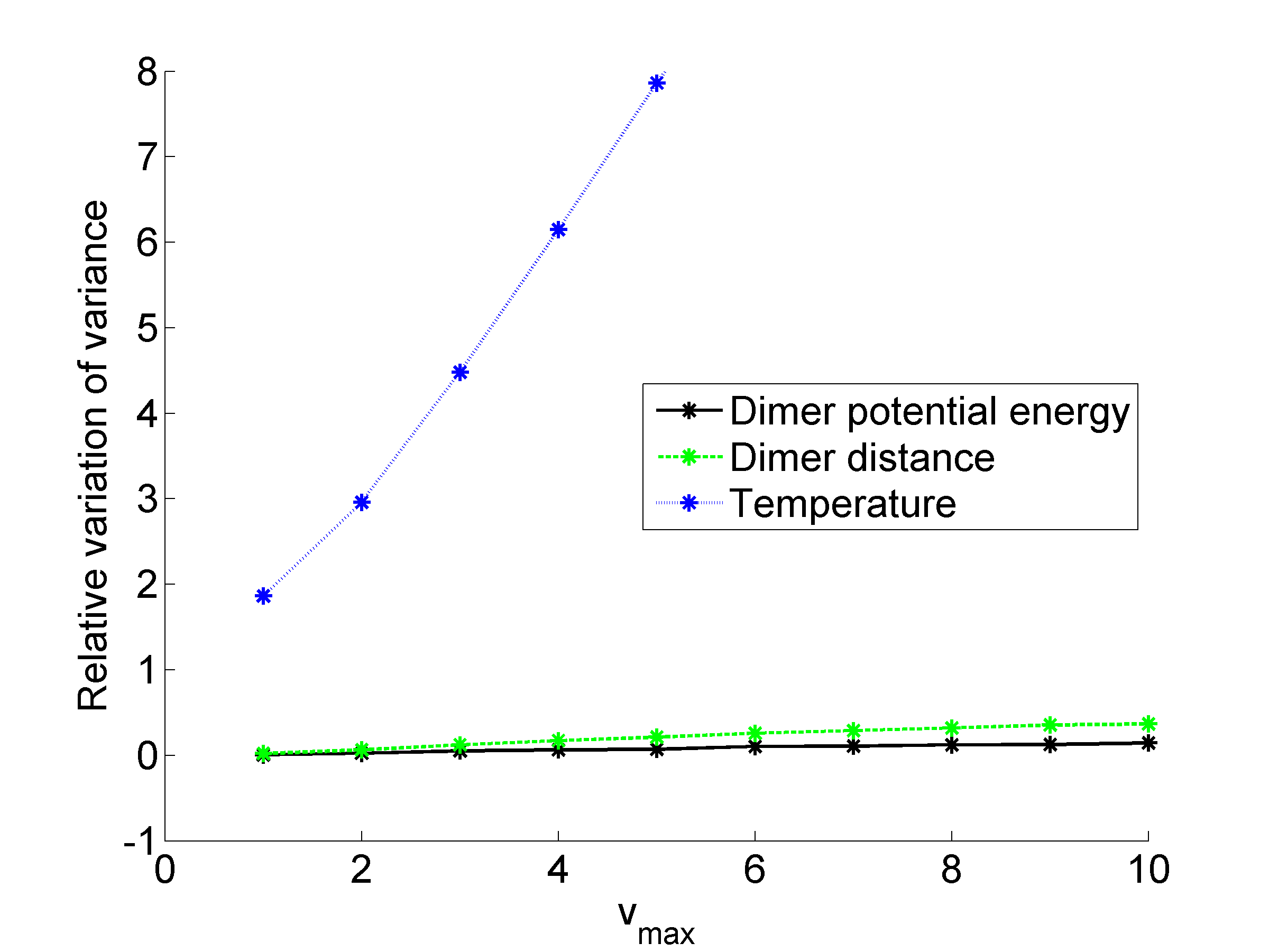}
			\caption{\textit{Relative deviation of the asymptotic variance from the variance in the standard dynamics.} We considered a system of described in Section \ref{section numerical examples}. We measured the variance of the dimer distance $A_D$ (green), the dimer potential $A_{DW}$ (black) and the temperature $\mathcal{T}$ (blue) in various parametrization of the AR dynamics. We plotted the relative deviation of the variance from the variance in the standard dynamics for parameters $\ef$ with $\delta=0.8$. Note that the variance is more perturbed with respect to the standard simulation for $\mathcal{T}$ than for the other two observables, which depend only on the dimer particles that are not restrained.}
	\label{fig:varianceAllBig}
	\end{figure}

Figure \ref{fig:varianceAllBig} plots the estimated relative variation of the variance of three observables as a function of the parameters $\ef$ for $\er=\delta\ef$ with $\delta= 0.5$
.   Recall that only the solvent particles are restrained and therefore the variance of an observable as $\mathcal{T}$, which depends directly on these degrees of freedom, is more perturbed from the variance in the standard case than the variance of an observable depending on particles following the standard dynamics (the dimer). We confirm the results showed in \cite{trstanova2015errorAnalysisOfMLD}: the variance of $\mathcal {T}$ is modified more drastically than the variance of observables measured on the dimer with growing $\ef$.

\begin{figure}[h]
	\centering
			\includegraphics[width=0.80\textwidth]{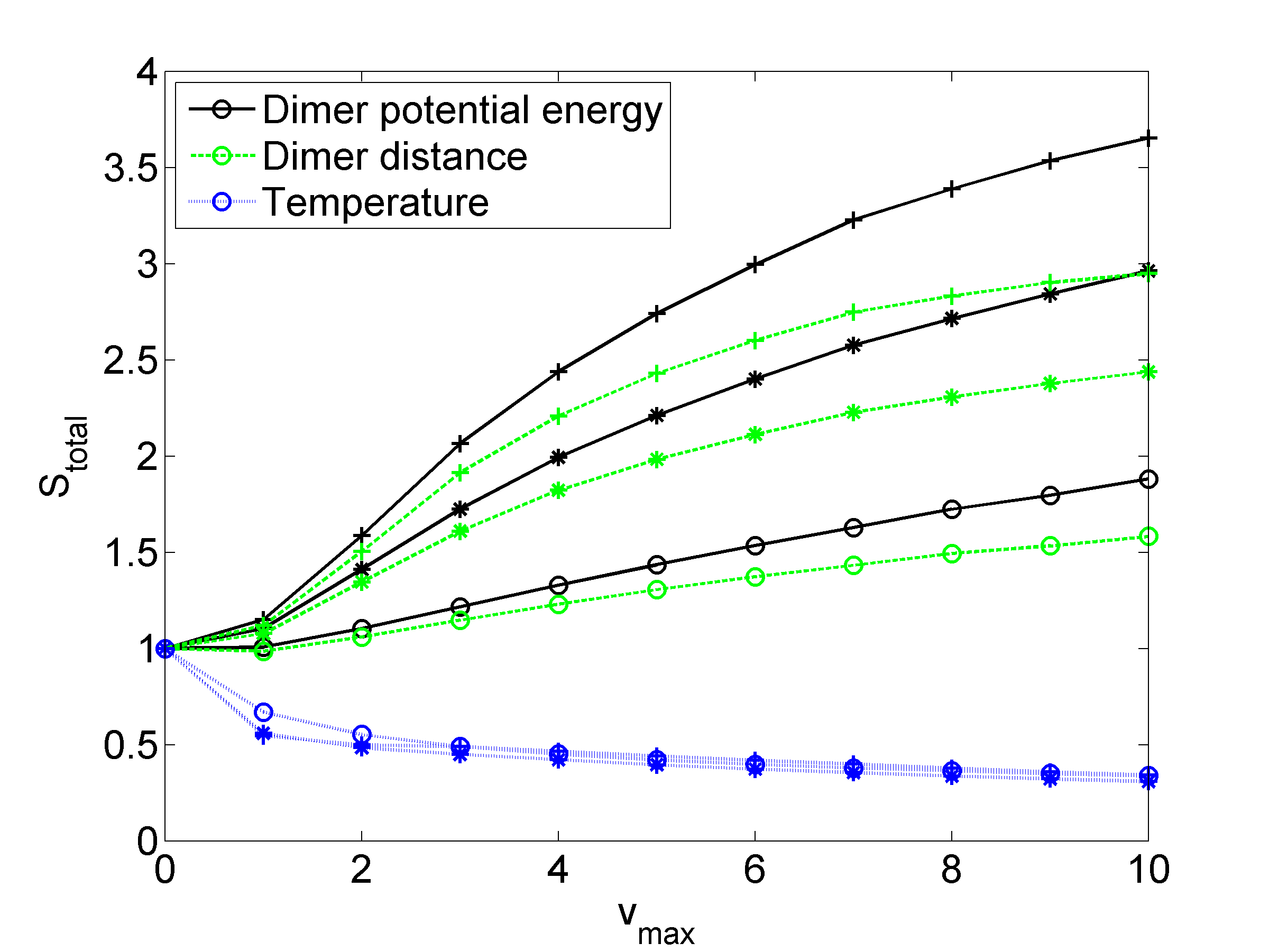}
			\caption{\textit{Numerical estimation of the total speed-up depending on the parameters ratio.} We estimated the total speed-up $S_{\rm total}$ given by \eqref{total speed-up} for system described in Section \ref{section numerical examples}. The algorithmic speed-up  as well as the variance (see Figure \ref{fig:varianceAllBig}) were obtained directly from the measurements in the simulation, \textit{i.e.} not from the analytical formula. On the plot the results are showed for three observables (various color and line styles). The various markers correspond to different value of $\delta$: the marker "open circles" corresponds to $\delta=0.5$, the marker "star" corresponds to $\delta=0.8$ and the marker "plus sign" corresponds to $\delta=0.9$, such that  $\delta=\er/\ef$. The bigger is $\er=\delta\ef$, the more particles are restrained which implies a bigger algorithmic speed-up. The bigger is $\delta$, the less perturbed is the variance and hence the better is $S_{\rm total}$.}
	\label{fig:total_speedUpBig}
	\end{figure}

Finally, combining the algorithmic speed-up $S_{\rm a}(\er,\ef)$ with the variance $\sigma^2(\er,\ef)$, we estimate the total speed-up according to \eqref{total speed-up}. This is depicted on Figure \ref{fig:total_speedUpBig}. We again consider $\delta_k\in \left\{0.5,0.8,0.9\right\}$ 
 in order to demonstrate the impact of the gap between the parameters $\delta=\er/\ef$ on the total speed-up $S_{\rm total}$: the smaller the gap, the larger $S_{\rm total}$ becomes. Also, it holds that $S_{\rm total}>1$  for the dimer observables only (up to $4$), and not for the global observable $\mathcal{T}$, for which the relative deviation of the variance dominates the algorithmic speed-up. 
This supports the idea that we can speed-up the computation of macroscopic properties that depend on unrestrained degrees of freedom, i.e. those of the dimer in this example.

It is easy to compute the algorithmic speed-up $S_{\rm a}$. The problematic part is the determination of the sensitivity of the observable on the restraining parameters (see again Figure \ref{fig:varianceAllBig}). Since the variance can be approximated by a linear function of the restraining parameters at least locally, we can compute the slopes $c_{\ef}$ such that\footnote{Note that, in this linear approximation, we consider a fixed ratio $\delta $ such that $\er=\delta \ef$.} $\sigma^2(\er,\ef)\approx\sigma^2(\er,\ef)+c_{\ef}\ef$ from three AR simulations with parameters $(\er^1,\ef^1),(\er^1,\ef^2),(\er^2,\ef^1)$. More precisely, this allows us to approximate the total speed-up as
	\begin{equation}
	S_{\rm{total}}(\er,\ef)\approx S_{{\rm a}}(\er,\ef)\frac{1}{1+\frac{c_{\ef}}{\sigma^2(0,0)}\ef}\,.
	\label{eq: rel dev variance num}
	\end{equation}
	We choose $\left(\er,\ef\right)\in \left\{(3,6),(3,7),(2,6)\right\}$ and we estimate the slope $c_{\ef}$. Table \ref{table variance three points} shows the comparison with the slope directly obtained from simulations with fixed $\delta=0.9$ and $\delta=0.8$ for $\ef\in \left[1,10\right]$ (see Figure \ref{fig:varianceAllBig}).
	\begin{table}[hp]
		\centering
			\begin{tabular}{|l||l|l|}
			\hline
			  $c_{\ef}/\sigma^2_{A_i}(0,0)$& $\delta=0.9$:  3 points & $\delta=0.9$: $\ef\in \left[0, 10\right]$ \\
			\hline
					$A_{DW}$ & $ 0.017863     $ &  $   0.016452     $ \\
					\hline
					$A_D $ &   $0.043855  $ & $  0.042729   $ \\
					\hline 					
					\hline
					& $\delta=0.8$: 3 points & $\delta=0.8$: $\ef\in \left[0, 10\right]$ \\
					\hline
					\hline
			$A_{DW}$	&$0.016426$ & $ 0.015028     $\\
				\hline
			$A_D $	& $0.039587$ & $0.039787$ \\
				\hline
			\end{tabular}
								\caption{Comparison of the relative deviation of the variance, obtained from linear interpolation of three points and the interpolation of values obtained from many simulations.}
								\label{table variance three points}
	\end{table}	
This confirms that it is possible to capture the quantitative behavior of the relative slope $c_{\ef}/\sigma^2_{\vp}(0,0)$ from only three AR simulations. Note that the same approach could be used to determine the behavior of the variance as a function of different temperatures by measuring the variance only at a few points.
	
\begin{figure}[h]
	\centering
			\includegraphics[width=0.80\textwidth]{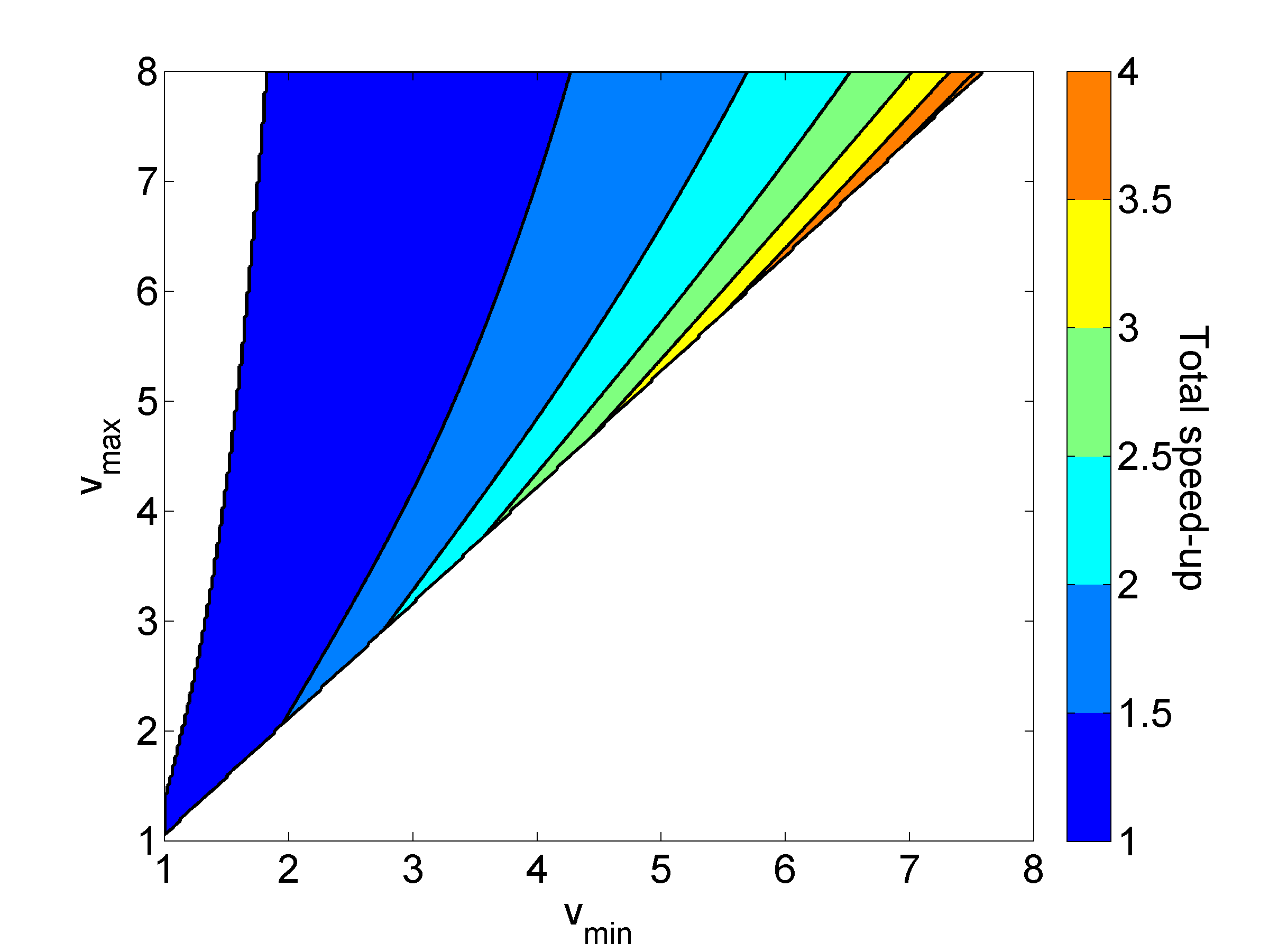}
			\caption{\textit{Analytical estimation of the total speed-up. }We estimated the expected total speed-up $S_{\rm total}$ for the observable dimer distance $A_D$ with respect to parameters $\er$ and $\ef$ ($\ef\leq 0.95\ef$). The variance was estimated from three points as a linear function of $\er$ and $\ef$ and we used the analytical estimation of $S_{\rm a}$ according to \eqref{eq: speed up newton nl}. Only $S_{\rm total}>1$ is plotted.}
	\label{fig:total_speedUp_analytical}
	\end{figure}

We have obtained an estimation of the variance $\sigma^2(\er,\ef)$. This allows us to express the total speed-up $S_{\rm total}$ as a function of $\er$ and $\ef$, which is depicted on Figure \ref{fig:total_speedUp_analytical} for $A_D$.
\begin{remark}
It would be possible to push the parameters in Figure \ref{fig:total_speedUp_analytical} in order to achieve a higher speed-up, up to the moment when the variance increase would start countering the algorithmic speed-up. Since the total speed-up depends on the simulated system and a concrete observable function, it does not make much sense to try and find the limit for our toy model. Moreover, for large parameters values, it is difficult to converge the quadratures in \eqref{Z_p er ef} and computationally too expensive to obtain the estimates numerically in the sense of Figure \ref{fig:total_speedUp_analytical}. Nevertheless, we believe Figure \ref{fig:total_speedUp_analytical} provides a good understanding of the qualitative behavior of the total speed-up.
\end{remark}
	
\section{Conclusions and future work}

We have analyzed the speed-up achievable by AR Langevin dynamics in the estimation of thermodynamics properties by time averages, in particular as a function of the restraining parameters $\er$ and $\ef$. The final formula consists of two parts: the algorithmic speed-up and the modification of variance. The approach proposed in this work allows us to choose the parameters of the method. The theoretical results are confirmed by numerical experiments. We expect that even higher total speed-ups can be achieved when the complexity of the incremental force update algorithm is improved, for instance by avoiding the double calculation of inter-particle forces (in the add and subtract steps).

The stability of the AR dynamics can be rigorously analyzed and improved by introducing a Metropolis-Hasting step, which is the purpose of future work (see \cite{NumericsInPreparation}).
Furthermore, the AR dynamics can be extended to simulations under different temperatures, and hence combined with other methods, such as parallel replica exchange (\cite{sugita1999replica}). Another natural extension is to explore the dynamical properties of the modified dynamics, which is not obvious at first sight.

\subsection*{Acknowledgements}
 Stephane Redon and Zofia Trstanova gratefully acknowledge funding from the European Research Council through the ERC Starting Grant n. 307629. We also thank Gabriel Stoltz and Svetlana Artemova for helpful advice.






\subsection*{Bibliography}

 \bibliographystyle{elsarticle-harv}
\end{document}